\documentclass[nonblindrev]{workingpaper}

\OneAndAHalfSpacedXII 

\usepackage{natbib}
 \bibpunct[, ]{(}{)}{,}{a}{}{,}%
 \def\bibfont{\small}%
\TheoremsNumberedThrough     

\EquationsNumberedThrough    

\usepackage{booktabs}
\usepackage{bm}
\newcommand\independent{\protect\mathpalette{\protect\independenT}{\perp}}
\def\independenT#1#2{\mathrel{\rlap{$#1#2$}\mkern2mu{#1#2}}}
\usepackage[ruled,vlined]{algorithm2e}
\usepackage{accents}

\usepackage{tikz,pgfplots}
\usepackage{caption,subcaption}
\usepackage{comment}
\usepackage{xfrac}

\usepackage{placeins}

\usepackage[hidelinks]{hyperref}
\begin{document}

\RUNAUTHOR{Pauphilet}

\RUNTITLE{Robust and Heterogenous Odds Ratio}

\TITLE{Robust and Heterogenous Odds Ratio: \\ Estimating Price Sensitivity for Unbought Items}

\ARTICLEAUTHORS{%
\AUTHOR{Jean Pauphilet}
\AFF{London Business School, London, UK, \EMAIL{jpauphilet@london.edu}}
} 

\ABSTRACT{%
{\it  Problem definition:} Mining for heterogeneous responses to an intervention is a crucial step for data-driven operations, for instance to personalize treatment or pricing.
We investigate how to estimate price sensitivity from transaction-level data. In causal inference terms, we estimate heterogeneous treatment effects when (a) the response to treatment (here, whether a customer buys a product) is binary, and (b) treatment assignments are partially observed (here, full information is only available for purchased items). 
{\it  Methodology/Results:} We propose a recursive partitioning procedure to estimate heterogeneous odds ratio, a widely used measure of treatment effect in medicine and social sciences. We integrate an adversarial imputation step to allow for robust estimation even in presence of partially observed treatment assignments. We validate our methodology on synthetic data and apply it to three case studies from political science, medicine, and revenue management.
{\it  Managerial Implications:} Our robust heterogeneous odds ratio estimation method is a simple and intuitive tool to quantify heterogeneity in patients or customers and personalize interventions, while lifting a central limitation in many revenue management data. 
}%



\maketitle

%


\section{Introduction}
Pricing and discount strategies are one of the most crucial decisions faced by (r)etailers. Given the amount of information available on every customer and shopping experience, merchants have a growing opportunity to personalize both of them. In particular, granular data at a customer or product level enables quantitative analysis of the impact of pricing and discounts on purchase. 
Motivated by transaction-level data 
made available as part of the M\&SOM 2020 data-driven challenge \citep{shen2019jd}, we propose a new methodology to conduct such an analysis. 
In doing so, we lift two methodological barriers: First, we propose a new partitioning procedure to estimate heterogeneous odds ratio, a popular metric to measure treatment effect in presence of binary responses. Secondly, we propose an adversarial data imputation method to derive robust findings despite the absence of full information on unpurchased items, a major limitation in revenue management applications involving transaction data (e.g., purchase panel, receipt, or loyalty data).

\subsection{A causal inference framework} \label{ssec:intro.causal}
We adopt terminology and notations from the potential outcome framework \citep[][Chapter 2]{imbens2015causal} and characterize observations by three quantities: $T \in \{0,1\}$ is the treatment assignment variable, such that $T=1$ (resp. $T=0$) for the treated (resp. control) individuals. $Y(T)$ captures the response to treatment $T$. Finally, $\bm{X}$ denotes a vector of side information or any observed covariates (e.g., gender or age). In this work, we focus on binary treatments $T \in \{0,1\}$ (we will relax this assumption in Section \ref{ssec:prescriptive}) 
and binary response to treatment $Y \in \{0,1\}$. The latter assumption holds in medicine application for instance, where ``response to treatment'' is defined as the occurence of adverse events such as death. 
Binary responses to treatment are also common in pricing applications (purchase decision), economics, and social sciences. 

Different measures for the effect of treatment $T$ on outcome $Y$ have been proposed. For instance, one can estimate the average response to treatment (resp. control)  $\mathbb{E}[Y(1)]$ (resp. $\mathbb{E}[Y(0)]$), or
the average treatment effect defined as $ATE = \mathbb{E}[Y(1) - Y(0)]$. Conceptually, these quantities summarize and compare the distributions of the two random variables $Y(1)$ and $Y(0)$. Unfortunately, for each patient, one does not observe $Y(1)$ and $Y(0)$ simultaneously, but only the response to the assigned treatment, i.e., $Y(T) | T=t$ where $t \in \{0,1\}$. This constitutes a major selection bias and the central problem of causal inference. This selection bias is usually addressed by assuming {\it unconfoundedness}, i.e., independence of {potential outcomes} and treatment assignment conditionally on the observed covariates \citep{rosenbaum1983central} and conducting a sensitivity analysis to assess the robustness of the findings to confounding factors \citep{rosenbaum2014sensitivity}. 

With the widespread availability of patient-level data in the past decade, the interest of the community and industry shifted from average to conditional/personalized treatment effect estimation. Indeed, treatments can now be assigned at a fine-grained scale. Hence, the effect of the treatment on a particular patient category is more relevant than its average on the total population. Mathematically, interest shifted from the  population-wide random variables $Y(0)$ and $Y(1)$ to their conditional versions, $Y(0) | \bm{X}=\bm{x}$ and $Y(1)  | \bm{X}=\bm{x}$. Mining for patterns of individual-level differences in treatment effect is also referred to as heterogeneous treatment effect (HTE) estimation. 
In particular and relevant to our work, tree-based methods from supervised machine learning have been adapted to the problem of heterogeneous treatment effect using decision trees \citep{zeileis2008model,su2009subgroup,laber2015tree,athey2016recursive,tran2019learning,bertsimas2019optimal,lee2020robust} and random forests \citep{wager2018estimation,athey2019generalized,lee2020causal}. 
While estimation of heterogeneous treatment effects can inform personalized treatment, \citet{lakkaraju2017learning,kallus2017recursive} propose a one-step approach to learn personalized policies directly, without an intermediate estimation step.
\citet{kallus2018confounding} investigate the question of policy learning in presence of confounding factors.

Recursive partitioning approaches proceed by iteratively dividing the observations into subgroups so as to minimize some splitting criterion. Various splitting criteria have been investigated, ranging from impurity measures \citep{chan2004lotus,landwehr2005logistic} 
to coefficient instability \citep{zeileis2008model}, variance of the estimates \citep{athey2016recursive}, prediction accuracy \citep{bertsimas2019optimal,aouad2019market}, $t$-statistics \citep{su2009subgroup}, and treatment effectiveness \citep{bertsimas2019optimal,kallus2017recursive}. Despite using different splitting criteria, these approaches unanimously accept the conditional average treatment effect $CATE = \mathbb{E}[Y(1) - Y(0)| \bm{X}=\bm{x}]$ as the metric to assess the effect of a treatment. When dealing with binary outcomes $Y \in \{0,1\}$, however, odds ratio are a common and intuitive measure of a treatment effect as well. The odds ratio $R$ compares the odds of responding when treated, $\mathbb{P}\left({Y} = 1 | T=1 \right) / \mathbb{P}\left({Y} = 0 | T=1\right)$, with the odds of responding when not treated, $\mathbb{P}\left({Y} = 1 | T=0 \right) / \mathbb{P}\left({Y} = 0 | T=0 \right)$, i.e., 
\begin{align} \label{eqn:or}
    R := \dfrac{\mathbb{P}\left( {Y} = 1 | T = 1\right)}{\mathbb{P}\left( {Y} = 0 | T = 1\right)} \times \dfrac{\mathbb{P}\left( {Y} = 0 | T = 0 \right)}{\mathbb{P}\left( {Y} = 1 | T = 0\right)}.
\end{align}
Defined as such, the odds ratio is simply a measure of association between $Y$ and $T$ but can be interpreted as a measure of causal effect under unconfoundedness. 
If $R > 1$ (resp. $R < 1$), the treatment increases (resp. decreases) the likelihood of responding positively. On the contrary, $R =1$ indicates that treatment and response to treatment are independent. Because it captures a treatment effect relative to the baseline risk, odds ratio are a widely used measure for treatment effect estimation in both randomized and observational studies, especially in meta-analysis that compare results from disparate studies \citep{engels2000heterogeneity}. The odds ratio relates with (but differs from) the relative risk or risk ratio \citep[see][for a discussion on the relative merits of each metric]{davies1998can}. Compared to difference in responses, ratio of responses does not only capture the magnitude of the effect but also its robustness to unmeasured confounding \citep[see][for a notorious discussion regarding tobacco and lung cancer]{cornfield1959smoking}. In our opinion, two reasons can explain the lack of interest in estimating heterogeneous odds ratio.
First, odds ratios suffer from non-collapsibility, a generalization of Simpson's paradox, which can hinder their interpretability \citep[Section 4.3]{hernan2020causal}-- note, however, that risk ratios {are collapsible if there is no confounding}. We refer to 
\citet{greenland1999confounding} for a discussion on non-collapsibility and its relation to confounding. Second, because ratios are relative measures, some authors expect them to be less heterogeneous than risk differences, while others have suggested that they can be equally heterogeneous but that tests for their heterogeneity has lower power \citep{poole2015risk}. This question remains unanswered.

In this paper, we propose a recursive partitioning procedure to evaluate heterogeneous odds ratio. Our procedure relies on Cochran's $Q$-test for heterogeneity and could be applied to a wide variety of treatment effect definitions, including odds ratios, risk ratios and CATE. To the best of our knowledge, we are the first to consider the heterogeneous odds ratio estimation task, which is surprising given the popularity of odds ratio in medicine and social sciences, and the profusion of research on HTE estimation.

\subsection{Challenges arising from real-world revenue management data} \label{ssec:intro.jd}
In the realm of the 2020 M\&SOM data-driven challenge \citep{shen2019jd}, researchers were given access to transaction-level data from the JD.com's e-tailing activity.  This data comprises information about millions of transactions that occured on the platform over the month of March 2018. For instance, demographic information about the customer, product attributes, pricing, and discount data are available. To estimate the impact of discount (the treatment) on the purchase decision (the response to treatment), our heterogeneous odds ratio estimation method would be a perfect candidate. 

Unfortunately,  data collection is triggered by a transaction and, by definition, a transaction corresponds to a customer {\it buying} an item. Consequently, the set of observations only includes customers and products that responded positively, rendering treatment effect estimation impossible. Note that this issue is present in many real-world revenue management settings. To alleviate this problem, we enrich our data set with information about user sessions: for each user, we know the pages this user visited and the products they considered but did not buy. Since these products have not been purchased, we only have partial information about the potential transaction. In particular, we do not know whether these products were offered to the customer with a discount. In other words, their treatment assignment is unknown. To summarize, the initial data set suffers from strong selection bias since all observations $(\bm{x}_i, t_i, y_i(t_i))$ satisfy $y_i(t_i)=1$. As an answer, we enrich the data set with partial observations of products that have been seen by the user but not bought, $(\bm{x}_i, \verb|NA|, 0)$, hence converting a selection bias problem into a missing data problem.

Missing data is a long-experienced problem in applied statistics. In the context of statistical inference, \citet{rubin1976inference} introduced the \emph{Missing At Random} (MAR) assumption and proved that inference was possible under this assumption. However, MAR is not a viable assumption in our case. Indeed, MAR requires that the missingness indicator (here, $Y$) is independent from the missing value (here, $T$), conditioned on the other covariates (here, $\bm{X}$). In our case, assuming MAR would negate the existence of an individual treatment effect and is not an acceptable assumption. An alternative is needed. Hence, we develop a robust imputation method integrated within the heterogeneous odds ratio procedure to find robust estimates of the odds ratio in presence of partially observed treatment assignments, without relying on the MAR assumption. In other revenue management applications, out-of-stocks can also be considered as partially observed treatment assignment \citep{musalem2010structural}, where the treatment is the assortment offered to the customer. Among others, \citet{vulcano2012estimating} developed an EM algorithm to properly estimate a demand model under some structural assumptions.

\subsection{Contributions and structure}
 Our contributions can be summarized as follows: 
\begin{itemize}
    \item We develop a recursive partitioning procedure to detect heterogeneous effect of treatment. Based on Cochran's $Q$-statistics, we propose a splitting criterion that explicitly aims at maximizing heterogenity across the subgroups in our partition, while accounting for statistical significance and sample size. Our criterion requires minimal assumptions on the estimator used to quantify the effect of treatment, namely asymptotic normality. 
    \item We implement this procedure in the case of binary treatments and binary responses, using odds ratio as a measure of treatment effect. Our approach, that we refer to as {\it Heterogeneous Odds Ratio} (HOR) estimation, provides an interpretable partition of the feature space together with asymptotically valid confidence intervals. Compared to approaches based on conditional average treatment effect, HOR estimation readily applies to randomized control trials as well as observational studies, and naturally informs confounding-robust policy improvement. 
    From HOR estimation, we derive personalized treatment assignment strategies that maximize the odds of responding positively ($Y=1$). Under this prescriptive lens, we extend our framework to multiple treatments.
    \item We propose a robust optimization approach integrated in our recursive partitioning scheme to address the issue of partially observed treatment assignment. In practice, this situation emerged for instance when the original data set suffers from a strong selection bias and additional data with unknown treatment assignment is collected. Our approach relies on defining a set of plausible set of treatment assignments and considering adversarially assigned treatments.
    \item We validate our methodology on synthetic data and apply it to case studies from political science, medicine, and revenue management. In particular, we apply our methodology to evaluate the effect of discount on purchase using JD.com transaction-level data, and identify that the original price and the product type reveal strong heterogeneity in the effectiveness of discounts. 
\end{itemize}

\paragraph{Structure: } We present the HOR estimation procedure and its implication for policy recommendation in Section \ref{sec:hor}. Section \ref{sec:theory} derives adaptive concentration bounds for our HOR estimates and discusses extensions to random forest estimation. 
Section \ref{sec:robust} addresses the missing treatment assignment problem using random sampling and robust optimization. We validate our approach on synthetic data in Section \ref{sec:syn} and then apply it to a collection of examples from social science and medicine (\S \ref{sec:appli}), and revenue management (\S \ref{sec:jd}).
\paragraph{Notations: } Nonbold lowercase characters ($x$) denote scalars and bold lowercase characters ($\bm{x}$) vectors. Uppercase letters designate random variables, e.g., $X$ and $\bm{X}$ respectively denote a random scalar and vector. The symbol $\independent$ designates independent random variables.

\section{Policy Learning via Heterogeneous Odds Ratio} \label{sec:hor}
In this section, we propose a recursive partitioning procedure for estimating heterogeneous treatment effects in presence of binary response to treatment. 
We measure the effect of treatment in terms of odds ratio (\S \ref{ssec:hor.or}) and use Cochran's $Q$-statistics as a splitting criterion (\S \ref{ssec:hor.q}). The final procedure is described in Section \ref{ssec:hor.alg}. We discuss policy implications of odds ratio in Section \ref{ssec:risk} and the extension to multiple treatments in Section \ref{ssec:prescriptive}.

\subsection{Odds ratio: Estimation and confidence intervals} \label{ssec:hor.or}
From $n$ observations of pairs $(t_i, y_i)$, construct a $2\times 2$ contingency table (Table \ref{tab:2x2cont}). The odds ratio compares the odds of responding positively when treated, $N_{1,1}/N_{0,1}$, with the odds of responding positively without treatment, $N_{1,0}/N_{0,0}$. An estimate for the log-odds ratio $\log R$ and its variance are given by
\begin{align}
    \log \hat{r} &= \log({N_{1,1}}) - \log({N_{1,0}}) + \log({N_{0,0}}) - \log({N_{0,1}}), \label{eqn:logor} \\
    \hat{v} = \hat{\sigma}^2( \log \hat{r}) &= {\dfrac{1}{N_{1,1}} + \dfrac{1}{N_{1,0}} + \dfrac{1}{N_{0,1}} + \dfrac{1}{N_{0,0}}}. \label{eqn:logor.var}
\end{align}
\begin{table}
\footnotesize
\caption{Sample $2\times 2$ contingency table, where $N_{t,y} = \sum_{i} \bm{1}(T_i=t) \bm{1}(Y_i=y)$. }
\label{tab:2x2cont}
    \centering
    \begin{tabular}{c|cc}
        & $Y=0$ & $Y=1$  \\
        \midrule
        $T=0$ &  $N_{0,0}$ & $N_{0,1}$\\
        $T=1$ & $N_{1,0}$ & $N_{1,1}$
    \end{tabular}
\end{table}
By the large-sample normality of $\log \hat{r}$ \citep{breslow1981odds}, $\log \hat{r} \pm z_{\alpha/2} \,  \hat{\sigma}( \log \hat{r}) $, with $z_{\alpha/2}$ the $\alpha/2$ quantile of the standard distribution, is a Wald confidence interval for $\log R$ . Exponentiating its endpoints provides a confidence interval for $R$. This interval was first propsed by \citet{woolf1955estimating} and performs quite well compared to subsequent proposals \citep[see][for a comparison]{lawson2004small}. Its robustness for small samples can be improved by adding a $+1/2$ bias term to each cell value $N_{t,y}, y\in \{0,1\}, t\in\{0,1\}$ \citep{jb1956estimation,gart1967bias,gart1966alternative}. We use this small-sample correction in our implementation.

\subsection{Testing for heterogeneity} \label{ssec:hor.q}
We now consider $K$ disjoint subgroups and estimate the sample log-odds ratio $\log \hat{r}^{(k)}$ and its standard error $\hat{v}^{(k)}$ on each subgroup separately. Borrowing concepts from hypothesis testing, we define a criterion to measure heterogeneity between subgroups.

Let us illustrate the case with two subgroups, $K=2$. Since the observations are independent and the subgroups are disjoint, $\log \hat{r}^{(1)}$ and $\log \hat{r}^{(2)}$ are independent. Hence, $\log \hat{r}^{(1)} - \log \hat{r}^{(2)}$ is asymptotically normal with mean $\log {r}^{(1)} - \log {r}^{(2)}$ and variance $\hat{v}^{(1)} + \hat{v}^{(2)}$. We want to test whether ${r}^{(1)} = {r}^{(2)}$ ($H_0$). Under ($H_0$), $Z := \tfrac{\left( \log \Tilde{r}^{(1)} - \log \Tilde{r}^{(2)} \right)^2}{{\hat{v}^{(1)} + \hat{v}^{(2)}}}$ follows a chi-square distribution with one degree of freedom. In this case, $Z$ corresponds to McNemar's test statistics. Maximizing heterogeneity between subgroups can be achieved by maximizing the probability of ($H_0$) being rejected, i.e., minimizing
\begin{align*}
\mathbb{P} \left( Z \geq \dfrac{\left[ \log \hat{r}^{(1)} - \log \hat{r}^{(2)} \right]^2}{{\hat{v}^{(1)} + \hat{v}^{(2)}}} \right) = 1 - \chi_1^2\left( \dfrac{\left[ \log \hat{r}^{(1)} - \log \hat{r}^{(2)} \right]^2}{{\hat{v}^{(1)} + \hat{v}^{(2)}}} \right),
\end{align*}
where  $\chi_1^2$ denotes the cumulative distribution function of a chi-square random variable with one degree of freedom. 

This procedure extends to the case with $K$ subgroups and is referred to as Cochran's $Q$-test \citep{cochran1954combination}. We construct $K$ differences $\hat{\theta}^{(k)} - \bar{\theta}$. Here, we denote $\hat{\theta}^{(k)} := \log  \hat{r}^{(k)}$ for concision and define $\bar{\theta}$ as a weighted average of all $\hat{\theta}^{(k)}$'s: 
\begin{align*}
    \bar{\theta} = \dfrac{1}{\sum_{k=1}^K \hat{w}^{(k)}} \sum_{k=1}^K \hat{w}^{(k)} \hat{\theta}^{(k)}, \quad \mbox{with } \hat{w}^{(k)} = 1 / \hat{v}^{(k)}.
\end{align*}
Then, Cochran's $Q$ statistic is the weighted sum of the squared deviations of the individual effects from their average; that is
\begin{align} \label{eqn:qstat}
    Q := \sum_{k=1}^K \hat{w}^{(k)} (\hat{\theta}^{(k)} - \bar{\theta})^2.
\end{align}
Under mild assumptions and under the null hypothesis that $\theta^{(1)} = \dots = \theta^{(K)}$, $Q$ asymptotically follows a chi-square distribution with $K-1$ degrees of freedom. Hence, miminizing $1 - \chi_{K-1}^2\left(Q \right)$, or equivalently maximizing $Q$, yields greater heterogeneity. We refer to \citet{huedo2006assessing,kulinskaya2015accurate} for recent discussions on the use of the $Q$-test in comparing odds ratio.  

In the following section, we will use the $Q$-statistics as a measure of heterogeneity and construct a partition of the feature space to maximize it. Related to our work, \citet{su2009subgroup} use the $t$-statistics to measure heterogeneity in CATE within subgroups and build their partition. However, Student's $t$-test can only be applied to test homogeneity in sample means and cannot be used for comparing (odds) ratios. On this regard, using Cochran's $Q$-test is more general and can be used to assess homogeneity of any estimator, provided that the estimator's distribution is asymptotically normal with computable variance.

\subsection{A recursive partitioning procedure} \label{ssec:hor.alg}
Let $\Pi$ be a partition of the covariate space into $|\Pi|$ disjoint subgroups. Write $\ell_k$, $k=1,\dots,|\Pi|$, an element of $\Pi$. $\ell_k$ will equivalently be referred to as a leaf or a subgroup. By definition, for any $\bm{x}$, there exists a unique subgroup of $\Pi$ containing $\bm{x}$, denoted $\ell(\bm{x},\Pi)$.

Consider a dataset of $n$ observations $\mathcal{D}_n = \{(\bm{x}_i, t_i, y_i),\, i=1,\dots,n\}$ and a partition $\Pi$. 
In each subgroup, one can estimate the log-odds ratio \eqref{eqn:logor}, its variance \eqref{eqn:logor.var} and compute the $Q$-statistics \eqref{eqn:qstat}. Let us denote $Q(\Pi; \mathcal{D}_n)$ its value. The objective is to construct a partition that maximizes the $Q$-statistics, i.e., solve 
\begin{align*}
\max_{\Pi \text{partition}} \:  Q(\Pi; \mathcal{D}_n).
\end{align*}
Instead of solving the above maximization problem exactly over the entire set of partitions, we solve it recursively. The benefits of recursive partitioning are fourfold: First, the resulting partition has a natural tree structure, which is highly interpretable and easy to understand for the practitioner. Interpretability of models and explainability of decisions are indeed a growing concern of consumers and regulators \citep{goodman2017european}. In addition, trees constructed in a greedy fashion can be explained sequentially: the first split of the tree explains most of the heterogeneity, followed by the second split, and so on. In other words, depth orders the splits by decreasing importance. This is not the case of globally optimal trees.
Secondly, despite a vibrant corpus of works on formulating decision trees as mixed-integer optimization problems and solving them as such to provable optimality 
\citep[see, e.g.,][]{bertsimas2017optimal,gunluk2018optimal,zantedeschi2020learning,aglin2020learning,demirovic2020murtree,lin2020generalized}, 
greedy procedures remain the gold-standard for balancing accuracy, interpretability, and scalability.
Thirdly, recursive procedures can handle complex non-convex criteria like the $Q$-statistics. Finally, while controlling the $p$-value of Cochran's $Q$-test ensures that the odds ratio is not constant across all leaves, our recursive procedure will satisfy a stronger property, namely that no leaf can be further divided so as to detect heterogeneity in odds ratio (at a given significance level). From a hypothesis testing perspective, a greedy procedure will control for pairwise heterogeneity between sibling leaves (see Proposition \ref{prop:homogeneous} for a formal statement). 

We now describe our recursive partitioning procedure. Consider a partition $\Pi$ and a leaf $\ell$ of this partition. We investigate whether $\ell$ can be further divided. A split is defined as a coordinate $j$ and a threshold $t$. The split $(j,t)$  divides $\ell$ into two subgroups $\ell_- = \ell \cap \{\bm{x} : x_j < t \}$ and $\ell_+ = \ell \cap \{\bm{x} : x_j \geq t \}$. There are finitely many splits, so by explicit enumeration, one can find the one that maximizes the $Q$-statistics $Q(\{\ell_-,\ell_+ \},\mathcal{D}_n)$, denoted $(j^\star, t^\star)$. If the resulting $p$-value, $1 - \chi_1^2(Q(\{\ell^\star_-,\ell^\star_+ \},\mathcal{D}_n))$, exceeds a given threshold $p_{\max}$, then we do not split $\ell$. Otherwise, we remove $\ell$ from the partition, and add $\ell^\star_+$ and $\ell^\star_-$ instead. The algorithm terminates when no leaf in the current partition can be divided any further. Pseudo-code is given in the online appendix, Algorithm \ref{alg:rp}.

The recursive procedure described in Algorithm \ref{alg:rp} generates a partition $\Pi$ of the feature space that has a natural tree structure. The internal nodes of the tree correspond to the all the sub-groups $\ell$ considered and split by the algorithm (e.g., its root node is the entire space $\mathbb{R}^p$) , while its leaves or terminal nodes are exactly the elements of $\Pi$. With a slight abuse of notation, we will also denote $\Pi$ the tree associated with the construction of the partition $\Pi$. We define \emph{siblings} as two nodes in the tree $\Pi$ that share the same parent. Partitions produced by our algorithm (Algorithm \ref{alg:rp}) satisfy the following property:
\begin{proposition} \label{prop:homogeneous} 
Let $\Pi$ be the tree resulting from applying Algorithm \ref{alg:rp} on the training data $\mathcal{D}_n$. For any siblings $node_1$ and $node_2$, the hypothesis that $node_1$ and $node_2$ are homogeneous in terms of odds ratio is rejected by Cochran's $Q$-test on $\mathcal{D}_n$, at a level $p_{\max}$.
\end{proposition}

\subsection{Odds ratio as a treatment recommendation} \label{ssec:risk}
As mentioned in introduction, comparing the odds ratio to $1$ provides an indication on whether the treatment is effective or not. The main advantages of odds ratio as a measure of causal effect are its simplicity, its symmetry with respect to outcome definition, and the fact that it is insensitive to the treatment propensity and baseline effect, $\mathbb{P}(T=1)$  and $\mathbb{P}(Y=1|T=0)$ respectively. If we assume that $Y=1$ is a positive outcome (e.g., survival or purchase), $R>1$ implies that the treatment increases the odds of $Y$ being positive, hence is beneficial. Odds ratio are prominently used in situations where assigning a reward or value to each outcome is impossible (e.g., in medicine when the outcome is death).

Due to their non-collapsibility, odds ratios usually lack any interpretation either as the change in average odds or the average change in odds, contrary to risk ratios \citep{cummings2009relative}. For instance, in revenue management, the risk ratio bounds the {allowable} discount: 
Assume $p_1$ (resp. $p_0$) is the discounted (resp. original) price. Then, from a retailer's perspective, discounting makes sense as long as the expected revenue under discount, $p_1 {\mathbb{P}\left( {Y} = 1 | T = 1\right)}$, exceeds the baseline revenue $ p_0 {\mathbb{P}\left( {Y} = 1 | T = 0\right)}$, i.e., as long as the risk ratio ${\mathbb{P}\left( {Y} = 1 | T = 1\right)} / {\mathbb{P}\left( {Y} = 1 | T = 0\right)}$ is greater than ${p_0}/{p_1}$. 

The interpretation of odds ratio is not as straightforward. Define $\text{odds}(z) := z / (1-z)$ for any $z \in [0,1)$. The function $z \mapsto \text{odds}(z)$ is a non-linear, increasing, and convex transformation of $z$. From the definition of the odds ratio $R$ in \eqref{eqn:or}, we have
\begin{align*}
\mathbb{P}(Y=1|T=1) = \text{odds}^{-1}\left( R \, \times \, \text{odds}\left( \mathbb{P}(Y=1|T=0) \right)  \right),  
\end{align*}
a non-linear transformation of $\mathbb{P}(Y=1|T=0)$. Nonetheless, for rare events, i.e., $\mathbb{P}(Y=1) \to 0$, since $\text{odds}(z) \sim_{z \to 0} z$, odds ratios are equivalent to risk ratios and share the same interpretation.

Odds ratio also have a natural interpretation in the context of repeated experiments. For instance, consider a situation when a customer comes multiple times to a (virtual) store. Assume that visits are independent and identical. Then, the number of times a customer will come to the store before making a first purchase follows a geometric distribution over $\{0, 1, \dots\}$ and is equal to $\mathbb{P}(Y=0) / \mathbb{P}(Y=1)$ on average. In this setting, the odds ratio compares times to purchase with and without treatment. For example, if $R=2$, then treatment (e.g., discount) halves time to purchase. Maximizing the odds ratio is equivalent to minimizing time to purchase. Further assume that the decision-maker incurs a cost $c$ per visit and receives a reward $r$ as soon as $Y=1$. For example, if the decision-maker decides to treat, $c$ can correspond to the marginal cost of treatment. Then, the average reward is positive as long as $\text{odds}(\mathbb{P}(Y=1)) > \dfrac{c}{r-c}$. In other words, for a fixed reward $r$, the odds bound the cost per visit $c$ that the decision maker can afford.

Note that our splitting criterion only requires asymptotic normality of the estimates of the treatment effect, so Algorithm \ref{alg:rp} can be applied to other definitions of the treatment effect. For instance, it can be applied to the relative risk discussed earlier, ${\mathbb{P}\left( {Y} = 1 | T = 1\right)} / {\mathbb{P}\left( {Y} = 1 | T = 0\right)}$, which can be empirically estimated as (notations from Table \ref{tab:2x2cont}):
\begin{align*} 
    \dfrac{N_{1,1}}{N_{1,1}+N_{1,0}} \times \dfrac{N_{0,0}+N_{0,1}}{N_{0,0}}.
\end{align*}
 
As for odds ratio, unconfoundedness is needed to interpret risk ratios as a measure of causal effect and not association only. Furthermore, in observational studies, techniques like inverse propensity score weighting are needed for unbiased estimation. In these settings, the risk ratio between the responses to treatment, ${\mathbb{P}\left( {Y}(1) = 1 \right)}/{\mathbb{P}\left( {Y}(0) = 1 \right)}$ is equal to the risk ratio between the responses to assigned treatment, ${\mathbb{P}\left( {Y} = 1 | T = 1\right)}/{\mathbb{P}\left( {Y} = 1 | T = 0\right)}$, up to a multiplicative factor, $\mathbb{P}(T=1)/\mathbb{P}(T=0)$. Our recursive partitioning procedure can be used as a non-parametric estimation of these ``propensity odds''. Note that, for odds ratio, similar corrections are needed in observational studies. However, using inverse propensity weighting and our partition as a non-parametric model for propensity, the corrective terms would appear both on the numerator and denominator and Equations \eqref{eqn:logor}-\eqref{eqn:logor.var} would remain valid as is.

\subsection{Extension to multiple treatments} \label{ssec:prescriptive}
So far, we restricted our attention to a single treatment. We now consider the situation when one can chose between multiple treatments, i.e., $T$ can take values $0$ (control), $1,\dots,m$. For each treatment $T=t$, one can compute the odds ratio associated with this particular treatment:
\begin{align*}
    R(t) := \dfrac{\mathbb{P}\left( {Y} = 1 | T = t \right)}{\mathbb{P}\left( {Y} = 0 | T = t\right)} \times \dfrac{\mathbb{P}\left( {Y} = 0 | T = 0 \right)}{\mathbb{P}\left( {Y} = 1 | T = 0\right)}.
\end{align*}
Consequently, each leaf is characterized by an $m$-dimensional vector of odds ratios for the $m$ possible treatments. A naive way to extend our approach to multiple treatments would be to consider a vector extension of Cochran's $Q$-statistics to test the heterogeneity in the multi-variate response $(R(t))_{t=1,\dots,m}$ when running Algorithm \ref{alg:rp}. Unfortunately, these tests would have little power and might be of little relevance in practice. 

Instead, we adopt a more prescriptive and pragmatic approach: we associate each leaf with a scalar value corresponding to the best odds ratio among all treatments, namely $\max_{t={0},\dots,m} R(t)$. By doing so, a leaf in Algorithm \ref{alg:rp} will be further divided in two if it identifies two subgroups that have significantly different responses to the best treatment - but what the ``best'' treatment is might be different for each subgroup. {Observe that we consider control ($t=0$) as a candidate treatment with $R(0)=1$, so the tree can recommend not to treat if no treatment exhibits a positive effect.} \citet{kallus2017recursive} adopt a similar strategy for learning personalized policies for observational studies, when treatment effects are evaluated in terms of conditional average treatment effect.  

\section{Convergence Analysis and Application to Random Forests} \label{sec:theory}
In this section, we study the convergence of our recursive partitioning procedure for odds ratio estimation. We prove an adaptive concentration bound that states that, with high probability and simultaneously for any leaf $\ell$, the gap between the estimated odds ratio in leaf $\ell$ and the population-average odds ratio scales as  $\sqrt{\log (n) \log (p)/k}$. The main result is presented in Section \ref{ssec:t.bound}, after preliminary assumptions and notations (Section \ref{ssec:t.ass}). Section \ref{ssec:t.rf} in the online appendix discusses implications for random forest estimation.

\subsection{Notations and assumptions} \label{ssec:t.ass}
First, we impose that Algorithm \ref{alg:rp} generates $(\alpha,k)$-valid partitions: 
\begin{definition} A partition $\Pi$ is \emph{$(\alpha,k)$-valid} if it can be generated by a recursive partitioning scheme in which each child node contains at least a fraction $\alpha \in (0,0.5]$ of the data points in its parent node and each terminal node contains at least $k \in \mathbb{N}$ training examples. Given a dataset $\mathcal{D}_n$, we denote its set of  $(\alpha,k)$-valid partitions by $\mathcal{V}_{\alpha,k}(\mathcal{D}_n)$ (or simply $\mathcal{V}_{\alpha,k}$ when clear from the context). 
\end{definition}
This assumption can easily be enforced in practice by considering splits that leave child nodes with a minimum number of training samples. 

Our analysis relies on adaptive concentration inequalities for regression trees \citep{wager2015adaptive} and require a similar set of assumptions. First, a condition on the dependence of the individual coordinates:
\begin{assumption} \label{ass:dependent} (Weakly dependent features) The features $X \in [0,1]^p$ are distributed according to a density $f$ satisfying $\zeta^{-1} \leq f(x) \leq \zeta$, $\forall x\in [0,1]$, for some constant $\zeta \geq 1$.
\end{assumption}
Second, we require that the minimum leaf size $k$ grows reasonably fast as $n,p \rightarrow \infty$, i.e., 
\begin{assumption} \label{ass:minleaf} (Minimum leaf size) The minimum leaf-size $k$ grows with $n$ at a rate bounded from below by
\begin{align*}
\lim_{n \to \infty} \: \dfrac{\log(n) \max \left\{\log(p), \log \log(n) \right\}}{k} = 0.
\end{align*}
\end{assumption}

For any subgroup $\ell \subseteq \mathbb{R}^d$, we consider the $2 \times 2$ contingency tables (Table \ref{tab:2x2cont}) for this particular leaf and analyze the convergence of cell counts towards their population-average. In particular, for each cell $(t,y) \in \{0,1\}^2$, we compare
\begin{align*}
    \hat{\pi}^n_{t,y}(\ell) &= \dfrac{1}{| \{i : \bm{x}_i \in \ell \}| }\sum_{i : \bm{x}_i \in \ell} \bm{1}(t_i = t, y_i = y), \\
    \pi_{t,y}(\ell) &= \mathbb{P}(Y=y,T=t| \bm{X} \in \ell),
\end{align*}
and we define the corresponding log odds ratio: $\log(\hat{r}^n(\ell)) = \log(\hat{\pi}^n_{1,1}(\ell)) -  \log(\hat{\pi}^n_{1,0}(\ell)) -  \log(\hat{\pi}^n_{0,1}(\ell)) +  \log(\hat{\pi}^n_{0,0}(\ell))$; and similarly for $\log({r}(\ell))$.
\subsection{Adaptive concentration bounds} \label{ssec:t.bound}
Consider one tree $\Pi \in \mathcal{V}_{\alpha,k}$. For any leaf $\ell \in \Pi$, Hoeffding's inequality shows that $\hat{\pi}^n_{t,y}(\ell)$ concentrates around $\pi_{t,y}(\ell)$ at a rate $1 / \sqrt{k}$. Then, a standard Chernoff bound technique enables to bound the worst case deviation over all $\ell \in \Pi$ by
\begin{align*}
\mathbb{P} \left(  \sup_{\ell \in \Pi} \: \left| \hat{\pi}^n_{t,y}(\ell)  - {\pi}_{t,y}(\ell) \right| \geq \delta \right) \leq 2 \exp \left( \log\left(\tfrac{n}{k}\right) -2k\delta^2 \right),
\end{align*}
where we bound the total number of leaves in  $\Pi$ by $n/k$. In particular, it yields 
\begin{align} \label{eqn:nonadapt.concentration}
\lim_{n \to \infty} \: \mathbb{P} \left(  \sup_{\ell \in \Pi} \: \left| \hat{\pi}^n_{t,y}(\ell)  - {\pi}_{t,y}(\ell) \right| \geq \sqrt{\dfrac{1.1}{2 k} \log \left( \dfrac{n}{k} \right) } \right) = 0.
\end{align}
However, this bound is not adaptive in the sense that it is uniform over all leaves $\ell$ but for a particular partition $\Pi$. Our goal is to derive a bound that is also uniform over all potential partitions $\Pi \in \mathcal{V}_{\alpha,k}$, to bound the performance of any tree generated by Algorithm \ref{alg:rp}.

For the CART algorithm, \citet{breiman1984classification} derived a uniform convergence result for regression and classification trees. They first prove the uniform convergence of the empirical leaf distributions \citep[][Theorem 12.2]{breiman1984classification}. Then, under the assumption that the diameter of each leaf shrinks as $n \to \infty$ and under some regularity conditions, they conclude on the uniform converge of the empirical prediction error. Despite its prevalence \citep{gordon1978asymptotically,gordon1980consistent}, the vanishing leaf diameter assumption is admittedly strong. Our analysis relies on the recent work of \citet{wager2015adaptive}, who provide adaptive concentration bound for regression trees. As in \citet{breiman1984classification}, the key idea is to replace the supremum over all $\Pi \in \mathcal{V}_{\alpha,k}, \ell \in \Pi$ by a supremum over a finite set of regions of $\mathbb{R}^d$ whose size moderately increases with the dimension of the problem. Instead of relying on generic $\varepsilon$-nets from VC theory \citep{vapnik2015uniform} like \citet{breiman1984classification}, \citet{wager2015adaptive} exhibit a tailored set of approximating rectangles and leverage the fact that partitions are $(\alpha,k)$-valid together with Assumption \ref{ass:dependent}. Applied to our context, their result states as follows:
\begin{lemma} \label{lemma:wager}\citep[][Theorem 1]{wager2015adaptive} Suppose that we have a sequence of problems with parameters $(n, p, k)$ satisfying Assumptions \ref{ass:dependent}-\ref{ass:minleaf}. Then, sample averages of cell counts over all possible valid partitions concentrate around their expectations with high probability, i.e., for all $(t,y) \in \{0,1\}^2$:
\begin{align*}
\lim_{n,p,k \to \infty} \: \mathbb{P} \left(  \sup_{\Pi \in \mathcal{V}_{\alpha,k}, \ell \in \Pi} \: \left| \hat{\pi}^n_{t,y}(\ell)  - {\pi}_{t,y}(\ell) \right| \geq \dfrac{f(n,p,k)}{\sqrt k} \right) = 0,
\end{align*}
with $f(n,p,k) := \dfrac{9}{2} \sqrt{\dfrac{\log(n/k)\left( \log(pk) + 3 \log \log(n)\right)}{\log\left( (1-\alpha)^{-1} \right)}}$.
\end{lemma}
Notice that under Assumption \ref{ass:minleaf}, $f(n,p,k) / \sqrt{k} \to 0$, as $n,p,k \to \infty$. Moreover, in a moderately high-dimensional regime with $\liminf p/n > 0$, $f(n,p,k)$ scales as $\sqrt{\log(n)\log(p)}$.
Compared to the bound \eqref{eqn:nonadapt.concentration}, Lemma \ref{lemma:wager} is surprisingly strong. The ``cost''  of adaptively searching over all valid trees in high dimensions  only scales logarithmically with $pk$. 

Based on this lemma, we derive a concentration bound for our odds ratio estimates:
\begin{theorem} \label{thm:or}
Suppose that we have a sequence of problems with parameters $(n, p, k)$ satisfying Assumptions \ref{ass:dependent}-\ref{ass:minleaf}. Furthermore, assume that for all $(t,y) \in \{0,1\}^2$, there exists $\pi_0 > 0$ such that 
${\pi}_{y,t}(\ell) \geq \pi_0$, for all $\Pi \in \mathcal{V}_{\alpha,k}$, $\ell \in \Pi$.
Then, 
\begin{align*}
\lim_{n,p,k \to \infty} \: \mathbb{P} \left(  \sup_{\Pi \in \mathcal{V}_{\alpha,k}, \ell \in \Pi} \: \left| \log(\hat{r}^n(\ell)) - \log(r(\ell)) \right| \geq  \dfrac{8 f(n,p,k)}{\pi_0\sqrt k} \right) = 0,
\end{align*}
with $f(n,p,k) := \dfrac{9}{2} \sqrt{\dfrac{\log(n/k)\left( \log(pk) + 3 \log \log(n)\right)}{\log\left( (1-\alpha)^{-1} \right)}}$.
\end{theorem}
\begin{proof}{Proof} Denote $\mathcal{A}_n$ the event
\begin{align*}
    \mathcal{A}_n := \left\lbrace \forall (t,y) \in \{0,1\}^2,\ \sup_{\ell \in \mathcal{L}_n} \hat{\pi}^n_{y,t}(\ell) \geq \pi_0 / 2 \right\rbrace.
\end{align*}
According to Lemma \ref{lemma:wager}, $\mathbb{P}(\mathcal{A}_n) \xrightarrow[n \to \infty]{} 1$. Moreover, by concavity of the logarithm, for any $x,y \geq {\pi_0}/{2}$ we have $|\log(x) - \log(y)| \leq ({2}/{\pi_0})|x-y|$. Hence, on $\mathcal{A}_n$, for any $\ell \in \mathcal{L}_n$, we have 
\begin{align*}
|\log \hat{r}^n(\ell) - \log r(\ell)| &\leq \sum_{(t,y) \in \{0,1\}^2} \left| \log \left( {\hat{\pi}_{t,y}^n(\ell)} \right) - \log \left( {\pi_{t,y}(\ell)} \right) \right|
&\leq \dfrac{2}{\pi_0} \sum_{(t,y) \in \{0,1\}^2} \left|  \hat{\pi}_{t,y}^n(\ell) - \pi_{t,y}(\ell) \right|.
\end{align*}
Taking the supremum over $\ell$ on both sides yields, for any $\delta > 0$,
\begin{align*}
\mathbb{P} \left( \sup_{\Pi, \ell} \: \left| \log(\hat{r}^n(\ell)) - \log(r(\ell)) \right| \geq \delta \right) \leq 
\sum_{(t,y) \in \{0,1\}^2} \mathbb{P} \left( \sup_{\Pi, \ell} \left|  \hat{\pi}_{t,y}^n(\ell) - \pi_{t,y}(\ell) \right| \geq \dfrac{\pi_0 \delta}{8} \right) + \mathbb{P}(\mathcal{A}_n^c).
\end{align*}
Setting $\delta = 8 f(n,p,k)/\pi_0$ concludes the proof. \hfill \halmos
\end{proof}

Due to the discrete nature of the splits, decision trees can be notoriously unstable to perturbations in the data. To improve stability, \citet{breiman2001random} originally proposed to introduce randomization in the training process of each tree, train a collection of random trees, and aggregate them in a random forest. \citet{wager2015adaptive} leveraged their adaptive concentration bounds over trees to derive similar results for random forests. While a thorough analysis of heterogeneous odds ratio estimation via random forest is out of the scope of the present paper, we discuss the different alternatives available to extend our methodology to random forests in Appendix \ref{ssec:t.rf}.

\section{Adversarial Treatment Assignment Imputation} \label{sec:robust}
In this section, we develop a methodology to handle missing treatment assignment, which is a major concern in many pricing and revenue management applications. 
\subsection{Problem formulation} \label{ssec:rob.intro}
From a high-level perspective, our HTE estimation problem involves quantities of the form $\mathbb{E}[\theta(T,Y(T)) | \bm{X} = \bm{x}]$ for some function $\theta$ (e.g., cell counts in Table \ref{tab:2x2cont}) that we estimate empirically within each leaf. However, with transaction-level data, we do not observe the treatment $T$ whenever $Y=0$. In other words, we cannot estimate the distribution of $T, Y(T) | \bm{X} = \bm{x}$, rendering the estimation of cell counts, odds ratio, and $Q$-statistics impossible. 
As discussed in introduction, the Missing At Random assumption \citep{rubin1976inference} is not suited here for it implies that ${\mathbb{P}}(T=1|\bm{X} = \bm{x}) = {\mathbb{P}}(T=1|\bm{X} = \bm{x}, Y=1)$ and that the treatment has no effect.

This issue of unobserved treatment assignment is connected with imperfect compliance, i.e., situations when the treatment effectively delivered to each patient, $D$, differs from the assigned one, $T$. Even in randomized control trials where treatment assignment is random, imperfect compliance can bias estimation of the treatment effect. \citet{efron1991compliance} assume that compliance can be captured by an intensity of treatment delivered  (between 0 and 1) and estimate the ATE after making some structural assumptions on how the response depends on treatment intensity. For all-or-nothing compliance, \citet{manski1990nonparametric,balke1997bounds} derive valid upper and lower bounds on the ATE under imperfect compliance. \citet{angrist1996identification} propose to use an instrumental variable (e.g., the assigned treatment $T$) to properly estimate the effect of the received treatment $D$ \citep[see][for an application]{siddique2013partially}. 
\citet{bargagli2019heterogeneous,stoffi2020causal} incorporated this IV methodology into a recursive partitioning procedure for HTE estimation. However, these works assume that we observe the imperfect compliance, that is that we observe the assigned and the delivered treatment. On the contrary, in transaction-level data, there is no discrepancy between treatment assigned and delivered, but we do not observe the treatment for any observations that respond negatively ($Y=0$).

To circumvent this issue, we first make mild distributional assumptions on the distribution of $T,Y=0 | \bm{X} = \bm{x}$. These assumptions define a set of feasible distributions for $T,Y=0 | \bm{X} = \bm{x}$ and of feasible treatment assignments $\mathcal{T}$ (Section \ref{ssec:rob.set}). To incorporate uncertainty in treatment assignment into the estimation procedure, we adopt a conservative approach and replace all $Q$-statistics by their worst-case value over $\mathcal{T}$ (Section \ref{ssec:rob.rec}). Since finding the worst-case value over $\mathcal{T}$ is hard, we replace $\mathcal{T}$ by $N$ samples drawn uniformly from $\mathcal{T}$ and provide performance guarantee for this finite sample approximation (Section \ref{ssec:rob.sample}).

\subsection{Ambiguous treatment assignments} \label{ssec:rob.set}
In the case of binary treatment, the distribution of $T, Y=0 | \bm{X} = \bm{x}$ is fully parametrized by ${\mathbb{P}}(T=1|\bm{X} = \bm{x}, Y=0)$. However, in the JD.com data as in many revenue management data, only purchased items are observed. As a result, only the propensity score for purchased products, ${\mathbb{P}}(T=1|\bm{X} = \bm{x}, Y=1)$, can be estimated. We need to make some structural assumption to relate ${\mathbb{P}}(T=1|\bm{X} = \bm{x}, Y=1)$ and ${\mathbb{P}}(T=1|\bm{X} = \bm{x}, Y=0)$. 
\begin{assumption} \label{ass:mt} There exists a non-decreasing transformation $\varphi : [0,1] \rightarrow [0,1]$ such that
\begin{align*}
{\mathbb{P}}(T=1|\bm{X} = \bm{x}, Y=0) = \varphi \left( {\mathbb{P}}(T=1|\bm{X} = \bm{x}, Y=1) \right), \ \forall \bm{x}.
\end{align*}
\end{assumption}
Given two contexts $\bm{x}_1, \bm{x}_2$ such that $\mathbb{P}(T=1|\bm{X} = \bm{x}_1, Y=1) < \mathbb{P}(T=1|\bm{X} = \bm{x}_2, Y=1)$, Assumption \ref{ass:mt} implies that $\mathbb{P}(T=1|\bm{X} = \bm{x}_1, Y=0) \leq \mathbb{P}(T=1|\bm{X} = \bm{x}_2, Y=0)$, i.e., the pairwise comparison between  $\bm{x}_1$ and $\bm{x}_2$ remains unchanged. Hence, although ${\mathbb{P}}(T=1|\bm{X} = \bm{x}, Y=1)$ does not properly estimate the probability of treatment when $Y=0$, it can be used as a relevant ranking score among items. In short, we assume that discount assignments are different between the two groups but that the drivers of discount are similar. We believe this assumption is fairly general and is broadly applicable beyond revenue management applications. 

Assumption \ref{ass:mt} alone does not impose any constraint on the resulting odds ratio, which can still take any value between $0$ and $\infty$ (with $\varphi(z) = 0$ or $1$).
Problem-specific knowledge can then suggest furthers restrictions on the transformation $\varphi$, such as convexity, bounded Lipschitz constant, or decomposability over a given basis. For example, with our targeted discount application in mind, we can reasonably assume that discount does not negatively impact the probability of purchase, or equivalently:
\begin{assumption} \label{ass:better} For any context $\bm{x}$,
${\mathbb{P}}(T=1|\bm{X} = \bm{x}, Y=0) \leq  {\mathbb{P}}(T=1|\bm{X} = \bm{x}, Y=1)$.
\end{assumption}
Assumption \ref{ass:better} is equivalent to the monotonous treatment response assumption \citep{manski2003monotone}. Recall that, by symmetry, the odds ratio at $\bm{x}$ is equal to $\text{odds}(z) / \text{odds}(\varphi(z))$, with $z = \mathbb{P}(T=1|Y=1, \bm{X} = \bm{x})$. Consequently, since odds$(\cdot)$ is an increasing function, Assumption \ref{ass:mt} implies that the odds ratio is at least equal to $1$. The MAR mechanism (which corresponds $\varphi(z) = z$) satisfies Assumptions \ref{ass:mt}-\ref{ass:better} and leads to an odds ratio of precisely 1. Alternatively, the odds ratio can be arbitrarily large by taking $\varphi(z) \to 0$.

Together, Assumptions \ref{ass:mt}-\ref{ass:better} enforces that the quantities ${\mathbb{P}}(T=1|\bm{X} = \bm{x}_i, Y=0)$ on our training data belong to the set:
\begin{align*}
    \mathcal{T} := \left\{\bm{t} \in [0,1]^n \: : \: {t}_{\sigma(i)} \leq {t}_{\sigma(i+1)},\ t_i \leq {\mathbb{P}}(T=1|\bm{X} = \bm{x}_i, Y=1), \forall i \right\},
\end{align*}
where $\sigma$ is a permutation ordering the ${\mathbb{P}}(T=1|\bm{X} = \bm{x}_i, Y=1)$'s in increasing order.
Each group of constraints correspond to Assumption \ref{ass:mt} and \ref{ass:better} respectively. To construct the set $\mathcal{T}$ in practice, we fit a machine learning model to predict the propensity score on purchased items, $\hat{t}(\bm{x}) := \hat{\mathbb{P}}(T=1|\bm{X} = \bm{x})$. By applying the function $\hat t$ to unbought items, we estimate $\hat{t}_i := {\mathbb{P}}(T=1|\bm{X} = \bm{x}_i, Y=1)$ and construct $\mathcal{T}$.

\subsection{Robust recursive partitioning} \label{ssec:rob.rec}
A feasible treatment vector is obtained by sampling $n$ Bernouilli random variables with parameter $t_i$ for some $\bm{t} \in \mathcal{T}$.  As far as odds ratio estimation is concerned, the quantities $N_{t,y}$ need to measure the expected number of counts in each cell of the contingency Table \ref{tab:2x2cont} so, in the remainder of the section, we assume that treatment assignments can be fuzzy/non-binary for the unbought items and equal to $t_i$. Accordingly, $\mathcal{T}$ should now be interpreted as the set of possible treatment assignments for non-purchased items.

We adapt the HOR estimation procedure (Algorithm \ref{alg:rp}) to be robust against uncertainty in treatment assignments. The algorithm is described in pseudo-code in Algorithm \ref{alg:robrp}. We start by computing the treatment assignment vector that minimizes the overall effect of the treatment, i.e., we solve $\min_{\bm{t} \in \mathcal{T}} |1 - \hat{r}(\bm{t}, \bm{y})|$ where $\hat{r}(\bm{t}, \bm{y})$ denotes the empirical odds ratio evaluated on the entire data as in Section \ref{ssec:hor.or}. We then apply our recursive partitioning procedure depth-wise: we first split the root node (depth $0$) and then its left and right children (depth $1$), and so on. At the end each step, before growing the tree deeper, we update the treatment assignments so as minimize the global heterogeneity $Q(\Pi, \{(\bm{x}_i, t_i, y_i)\})$. All splits made during this step are then re-assessed: any split that no longer satisfies the splitting condition, $1 - \chi_1^2(Q(\{\ell^\star_-,\ell^\star_+ \},\mathcal{D}_n)) \leq p_{\max}$, is undone. In other words, at each step, we compute the most homogeneous data set according to the $Q$-statistics and the stratification from the current partition, and keep only the splits whose validity are robust to this adversarial treatment assignment. We denote this procedure Robust HOR, R-HOR in short. 
\begin{algorithm}
\footnotesize
\SetAlgoLined
\KwIn{$\{(\bm{x}_i, y_i, \hat{t}_i),\, i=1,\dots,n\}$}
\KwResult{Partition $\Pi$.}
 initialization: $\Pi \leftarrow \{\mathbb{R}^p \}$, $\bm{t} \in \argmin_{\mathcal{T}} |1 - \hat{r}(\bm{y}, \bm{t})| $, $\mathcal{D}_n \leftarrow \{(\bm{x}_i, t_i,y_i),  i=1,\dots,n\}$\;
 \For{$d \in \{1,\dots ,max\_depth\}$}{
 $\mathcal{L}_d \leftarrow \{\ell \in \Pi \: : \: depth(\ell) = d\}$ \tcp{Note: potentially empty} 
 \For{$\ell \in \mathcal{L}_d$}{
 $(j^\star, t^\star) \leftarrow \argmax_{(j,t)} \: Q(\{\ell_-,\ell_+\} ,\mathcal{D}_n)$ \;
  \If{$1 - \chi_1^2(Q(\{\ell^\star_-,\ell^\star_+ \},\mathcal{D}_n)) \leq p_{\max}$ \texttt{and} other criterion}
  {split $\ell$: $\Pi \leftarrow (\Pi \backslash \{\ell \}) \cup \{\ell_-,\ell_+ \}$}
 }
 $\displaystyle\bm{t} \in \argmin_{\bm{t} \in \mathcal{T}} Q\left(\Pi, \{(\bm{x}_i, t_i, y_i)\}\right)$\;
  \For{$\ell \in \mathcal{L}_d$}{
  \If{$1 - \chi_1^2(Q(\{\ell^\star_-,\ell^\star_+ \},\mathcal{D}_n)) > p_{\max}$}
  {unsplit $\ell$: $\Pi \leftarrow (\Pi \backslash \{\ell_-,\ell_+ \}) \cup \{\ell \}$}
 }
 }
\caption{Robust recursive partitioning algorithm for HOR (R-HOR)} \label{alg:robrp}
\end{algorithm}

The main shortcoming of our approach is that, whenever the set of feasible assignment $\mathcal{T}$ is large, the adversarial can generate treatment assignments that negate the revealed splits. Especially at the first step of the algorithm, when there is only one split to validate. In particular, we previously observed that Assumptions \ref{ass:mt}-\ref{ass:better} are not restrictive and allow the odds ratio to take any value between 1 and $+\infty$. We propose two solutions to mitigate this issue. First, we can introduce a reference vector $\bm{t}_0$ and solve instead
\begin{align} \label{eqn:rob.tradeoff}
   \min_{\bm{t} \in \mathcal{T}} \:  Q\left(\Pi, \{(\bm{x}_i, t_i, y_i)\}\right) \quad \mbox{s.t. } \| \bm{t} - \bm{t}_0 \|_1 \leq n \Gamma,
\end{align}
where $\Gamma$ is a parameter controlling the similarity between the treatment assignment $\bm{t}$ and the reference assignment $\bm{t}_0$. The vector $\bm{t}_0$ can capture prior beliefs on treatment assignments and ensures that the vectors $\bm{t}$ generated at each step of the algorithm remain similar. In our experiments, we take $\bm{t}_0$ equal to the treatment assignment vector obtained at the initialization step. In the robust optimization literature, $\Gamma$ is referred to as the budget of uncertainty \citep{bertsimas2004price} and controls the amount of uncertainty the decision maker wishes to be protected against.  We will explore the numerical impact of $\Gamma$ in Section \ref{sec:jd}. Second, in our implementation, we replace $\mathcal{T}$ by a finite-sample approximation. Doing so not only simplifies the minimization problem in \eqref{eqn:rob.tradeoff} but also excludes pathological cases while preserving theoretical guarantees, as described in the following section.

\subsection{Finite sample approximation} \label{ssec:rob.sample}
Algorithm \ref{alg:robrp} requires finding the worst case treatment assignment $\bm{t} \in \mathcal{T}$, when worst case is defined in terms of worst odds ratio (initialization) or minimal $Q$-statistics. 
Solving these optimization problems is intractable given the non-convexity of the objectives. Instead, we generate uniformly at random $N$ potential vectors $\bm{t}^{(j)} \in \mathcal{T}$ and take the worst case over the finite set $\mathcal{T}^N =\{\bm{t}^{(j)} , j=1,\dots,N \}$. Naturally, for any criterion $f$, we have 
$\displaystyle \max_{\bm{t} \in \mathcal{T}^N} \: f(\bm{t}) \leq \max_{\bm{t} \in \mathcal{T}} \: f(\bm{t}),$
and the larger the $N$, the better the approximation, as formalized in the following proposition:
\begin{proposition}\label{prop:sampling} Denote $z_N := \max_{\bm{t} \in \mathcal{T}^N} \: f(\bm{t})$. Given a violation probability $\varepsilon \in (0,1)$ and a confidence parameter $\beta \in (0,1)$, if  $N \geq \dfrac{2}{\varepsilon}\left(1 - \ln \beta \right)$, then, with probability at least $1-\beta$ (on the samples $t^{(j)}$), we have 
\begin{align*}
\mathbb{P}_{\bm{T}} \left( f(\bm{T}) \leq z_N  \right) \leq \varepsilon,
\end{align*}
where the probability $\mathbb{P}_{\bm{T}}( \cdot )$ corresponds to the sampling probability over $\mathcal{T}$
\end{proposition} 
\begin{proof}{Proof}
We consider a scalar decision variable $z$ and the robust constraint {$\displaystyle \max_{\bm{t} \in \mathcal{T}} \: f(\bm{t}) \leq z$}.
Under this lens, our finite sampling approach relates to the scenario approach in robust design optimization and follows immediately from Theorem 1 in \citet{campi2009scenario}. \hfill \halmos
\end{proof}

Proposition \ref{prop:sampling} states that, with probability $1-\beta$ over the samples used to construct $\mathcal{T}^N$, the inequality $\displaystyle f(\bm{T}) \leq \max_{\bm{t} \in \mathcal{T}^N} \: f(\bm{t})$ holds with high probability. Intuitively, $\beta$ corresponds to the probability of sampling a ``bad'' set $\mathcal{T}^N$ and $\varepsilon$ bounds the probability of $\bm{T}$ falling outside of {the convex hull of} $\mathcal{T}^N$. Note that the distribution of $\bm{T} \in \mathcal{T}$ in Proposition \ref{prop:sampling} needs to be the same one as the sampling distribution. 
In our numerical analysis, we sample uniformly from $\mathcal{T}$. Accordingly, {if} pathological cases (e.g., $\varphi(z) = 0$ or $z$) are of probability zero, sampling $N$ scenarios from $\mathcal{T}$ will exclude these pathological cases with high probability. 
Regarding the sampling of treatment assignment vectors $\bm{t}^{(j)} \in \mathcal{T}$, the hit-and-run algorithm \citep{smith1984efficient} provides a generic method to sample uniformly from any polyhedron. Given the simplicity of the constraints involved in the definition of $\mathcal{T}$, however, we implemented a simple constructive procedure described in Appendix \ref{sec:A.sampling}.

\section{Validation on Synthetic Data} \label{sec:syn}
In this section, we assess the validity of our approach on synthetic data. We should acknowledge the fact that whether heterogeneity between patients' is better captured by the ratio of responses instead of their difference depends on the problem at hand. As a result, synthetic examples where our procedure outperforms other methods from the HTE literature could be easily constructed, and vice versa. To provide a fair comparison, we focus our analysis on the quality of the $Q$-statistics as a splitting criterion. We apply Algorithm \ref{alg:rp} in a setting where responses are continuous and effect of treatment is evaluated in terms of average treatment effect. In this setting, methods from the HTE literature apply and we can compare the methods in their ability to detect heterogeneity.  

\paragraph{Data generation} We adopt a similar methodology as \citet{lee2020causal}.
For $j=1,\dots,p$, $X_j$ is sampled according to a Bernouilli distribution with parameter $p_j$, with $p_j \sim \mathcal{U}(0,1)$. We generate the response to treatment and control according to 
\begin{align*}
    Y(0) &= X_1 + \tfrac{1}{2} X_2 + X_3 + \kappa(X_1, X_2, X_3) +  \varepsilon,\quad \varepsilon \sim \mathcal{N}(0,1), \\
    Y(1) &= Y(0) + \tau_k(X_1,X_2),
\end{align*}
where $\kappa(x_1, x_2, x_3) = \exp(x_1 - x_2 x_3)$ creates non-linear heterogeneity in responses (both to treatment and control) while $\tau_k(x_1,x_2)$ captures the heterogeneous treatment effect and is one of the following:
\begin{itemize}
    \item 2-rule case: $\tau_k(x_1,x_2) = k$ if $x_1 = 0$ and $x_2 = 0$, $-k$ if $x_1 = 1$ and $x_2 = 1$, and $0$ otherwise. 
    \item 4-rule case: $\tau_k(x_1,x_2) = k$ if $x_1 = 0$ and $x_2 = 1$, $2k$ if $x_1 = 0$ and $x_2 = 0$, $-k$ if $x_1 = 1$ and $x_2 = 0$, $-2k$ if $x_1 = 1$ and $x_2 = 0$, and $1$ otherwise. 
\end{itemize}
The parameter $k$ controls the magnitude of the treatment effect. Finally, the treatment assignment variable $T$ is distributed according to the logit model $\mathbb{P}(T=1|\bm{X} = \bm{x}) = \text{logit}(-1 + x_1 - x_2 + x_3)$. 

\paragraph{Our method} Since responses are continuous, we apply our recursive partitioning procedure (Algorithm \ref{alg:rp}) to the case where the effect of treatment is measured in terms of conditional average treatment effect (CATE). In-sample estimates of the CATE and its variance are obtained within each leaf and used in the computation of the $Q$-statistics for each split. We fix a maximal depth of $10$ and $p_{\max} = 0.1$. 

\paragraph{Metrics} We sample $N=1,000$ observations, for different values of $k$ ranging between $0$ and $4$. Consider a partition $\Pi$ of the feature space. Each rule in the definition of $\kappa$ (e.g., $x_1 = 1$ and $x_2 = 1$) is associated with a set of leaves of $\Pi$, $\{\ell_r \in \Pi \: : \: r \in \mathcal{R} \}$, such that for any $r \in \mathcal{R}$, there is at least one observation $\bm{x}_i$ from the training data such that $\ell(\bm{x}_i, \Pi) = \ell_r$ and $\bm{x}_i$ satisfy the rule. Intuitively, $\{\ell_r \in \Pi \: : \: r \in \mathcal{R} \}$ should be understood as the smallest subset of leaves needed to cover all the observations satisfying the rule. With this definition, we measure the quality of rule detection in terms of two metrics: \emph{complexity}, i.e., the cardinality of $\mathcal{R}$, and \emph{purity}, i.e., the proportion of observations in $\{\ell_r \in \Pi \: : \: r \in \mathcal{R} \}$ that indeed satisfy the rule. We say that $\Pi$ perfectly detects a rule if it achieves a complexity as low as $1$ and a purity of $100\%$. For each partition, we report complexity and purity, averaged over all rules defining $\kappa$.

\paragraph{Benchmark} We compare our splitting criterion (Q-stat) to alternatives: the $t$-statistics of \citet{su2009subgroup} (t-stat), and the causal tree of \citet{athey2016recursive} (CT), as implemented in the \verb|R| package \verb|causalTree| \citep{athey2016recursive}. Following best practices, we prune the trees obtained in this fashion using a cross-validation criterion that can be different from the splitting one. We consider three cross-validation criteria, denoted \verb|TOT|, \verb|fit| and \verb|CT| in the \verb|causalTree| documentation. When reporting the complexity or the purity of t-stat/CT, we report the value obtained with the cross-validation criterion achieving the best performance, thus leading to optimistic performance of the benchmarks.   

\begin{figure}
\begin{subfigure}{\linewidth}
    \centering
    \begin{subfigure}{.45\linewidth}
    \includegraphics[width=\linewidth]{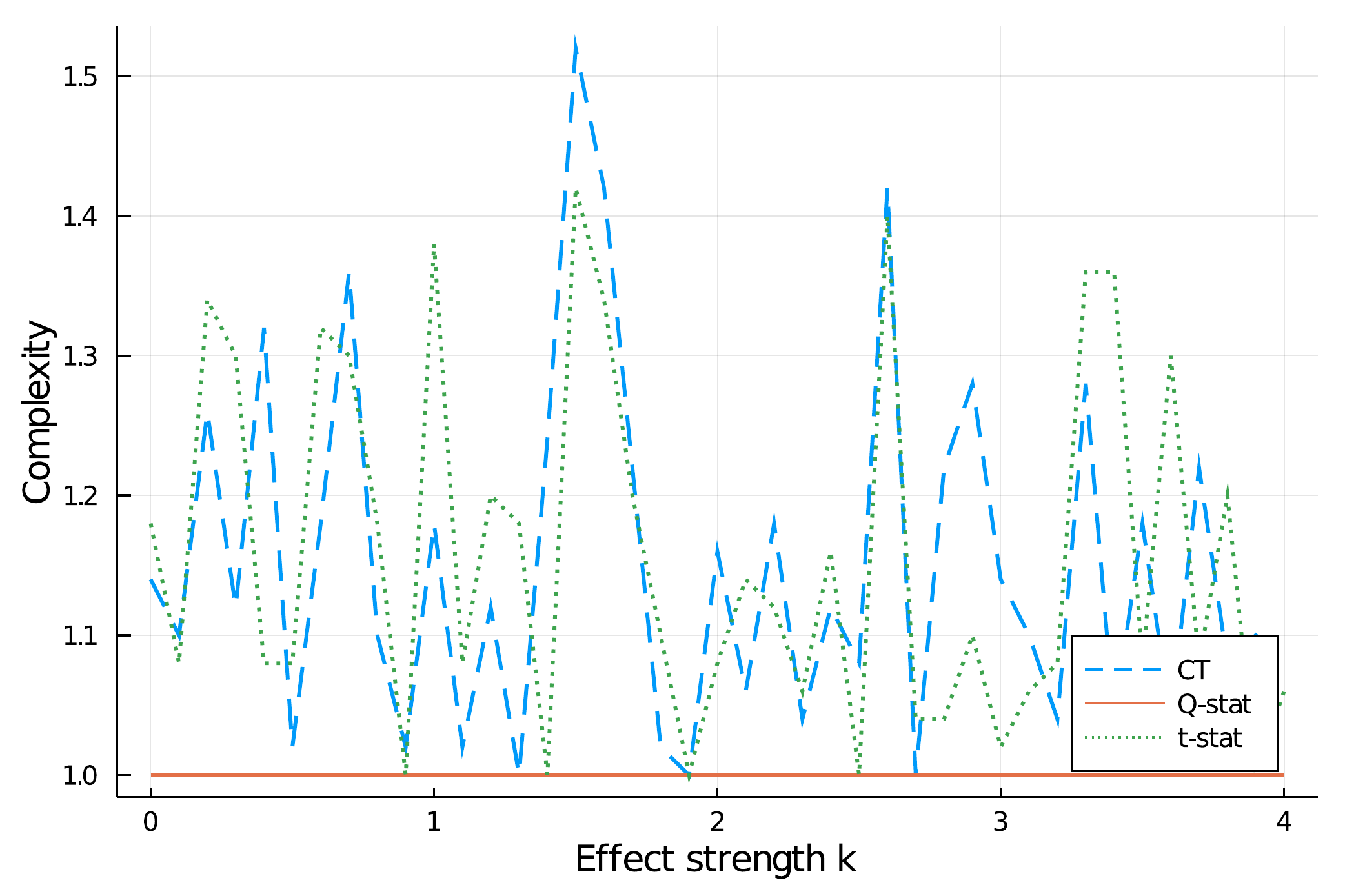}
    \end{subfigure} %
    \begin{subfigure}{.45\linewidth}
    \includegraphics[width=\linewidth]{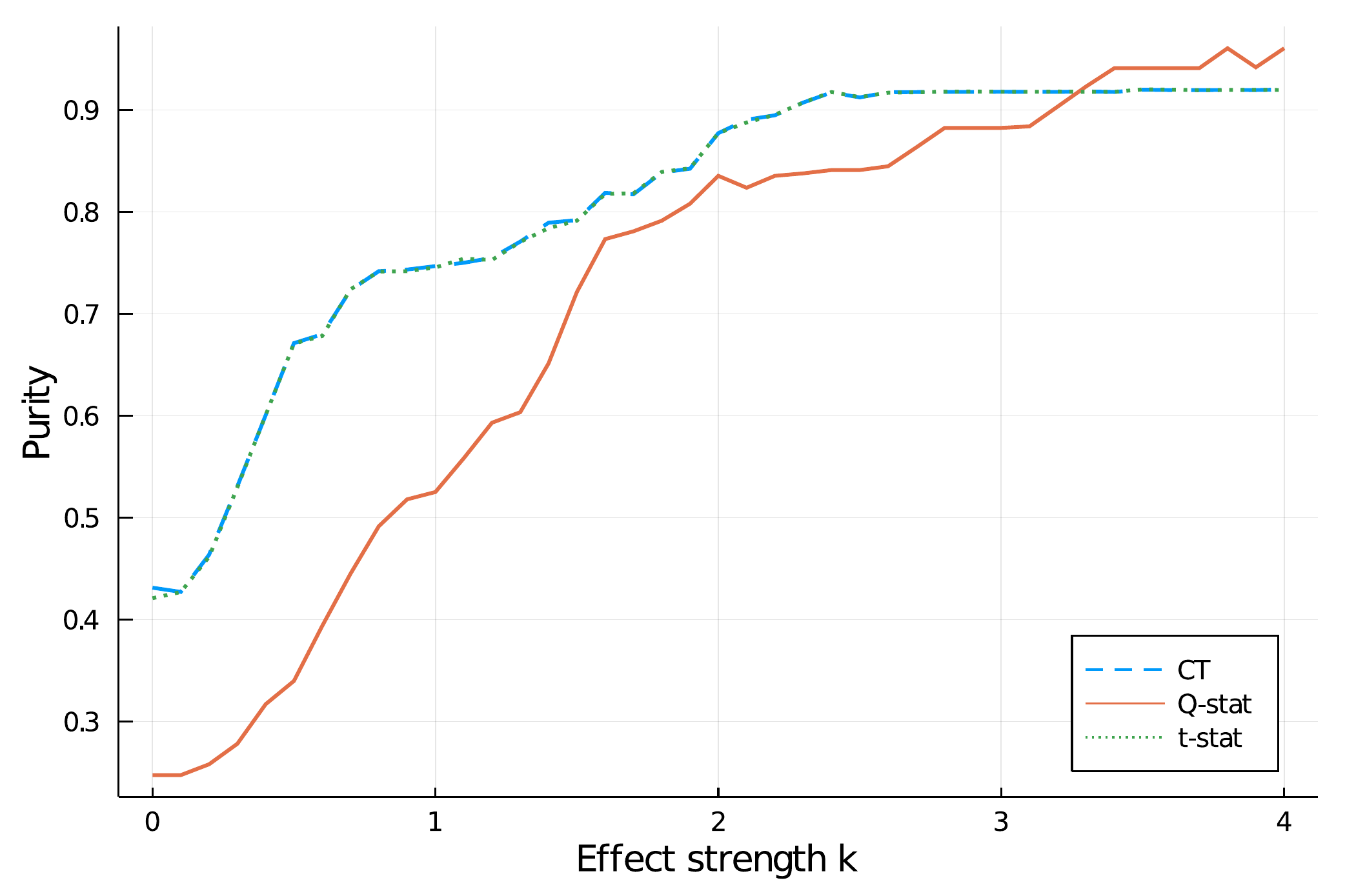}
    \end{subfigure}
    \caption{2-rule heterogeneous treatment effect}
    \label{fig:syn.2rule}
\end{subfigure}
\begin{subfigure}{\linewidth}
    \centering
    \begin{subfigure}{.45\linewidth}
    \includegraphics[width=\linewidth]{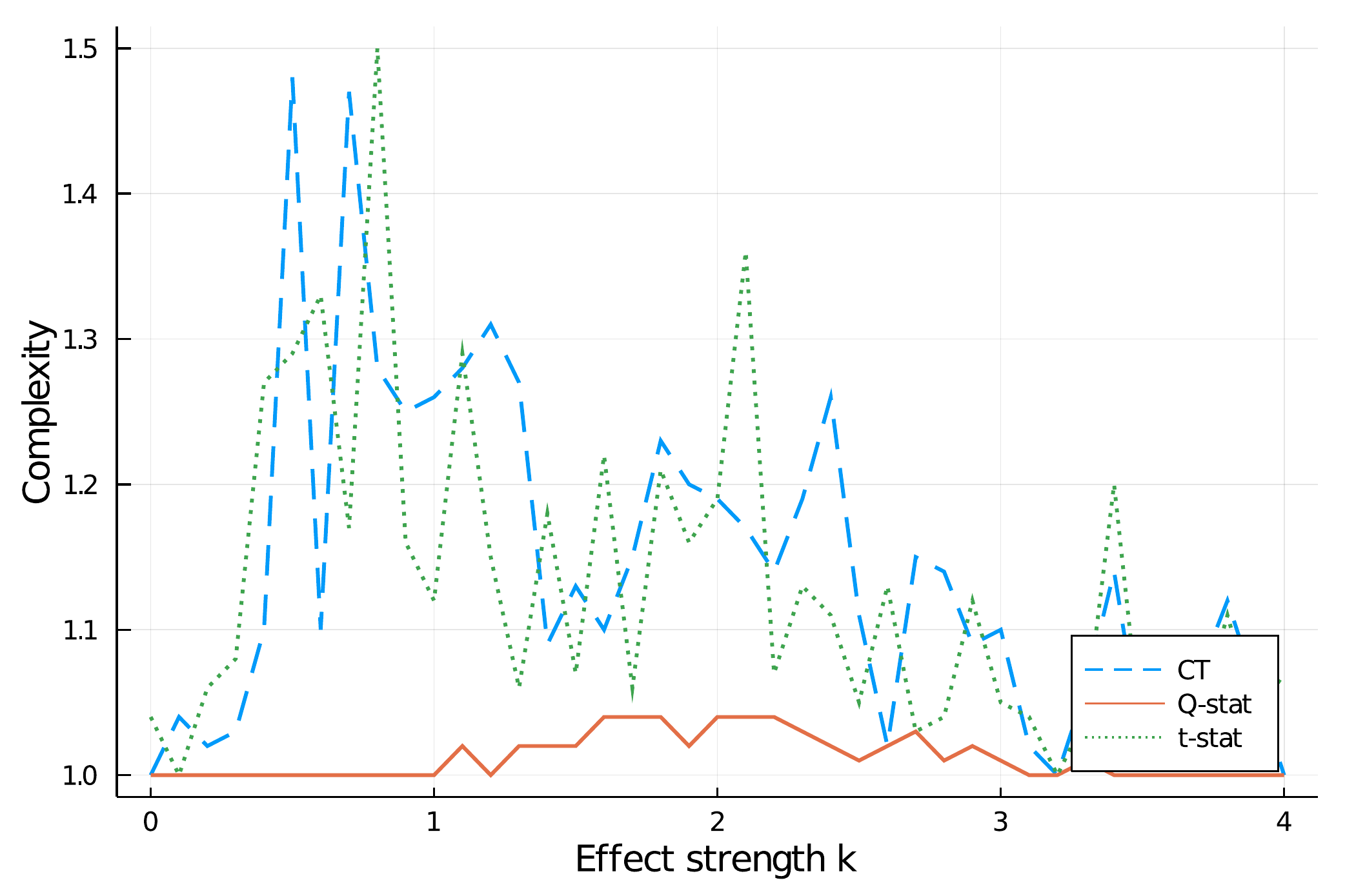}
    \end{subfigure} %
    \begin{subfigure}{.45\linewidth}
    \includegraphics[width=\linewidth]{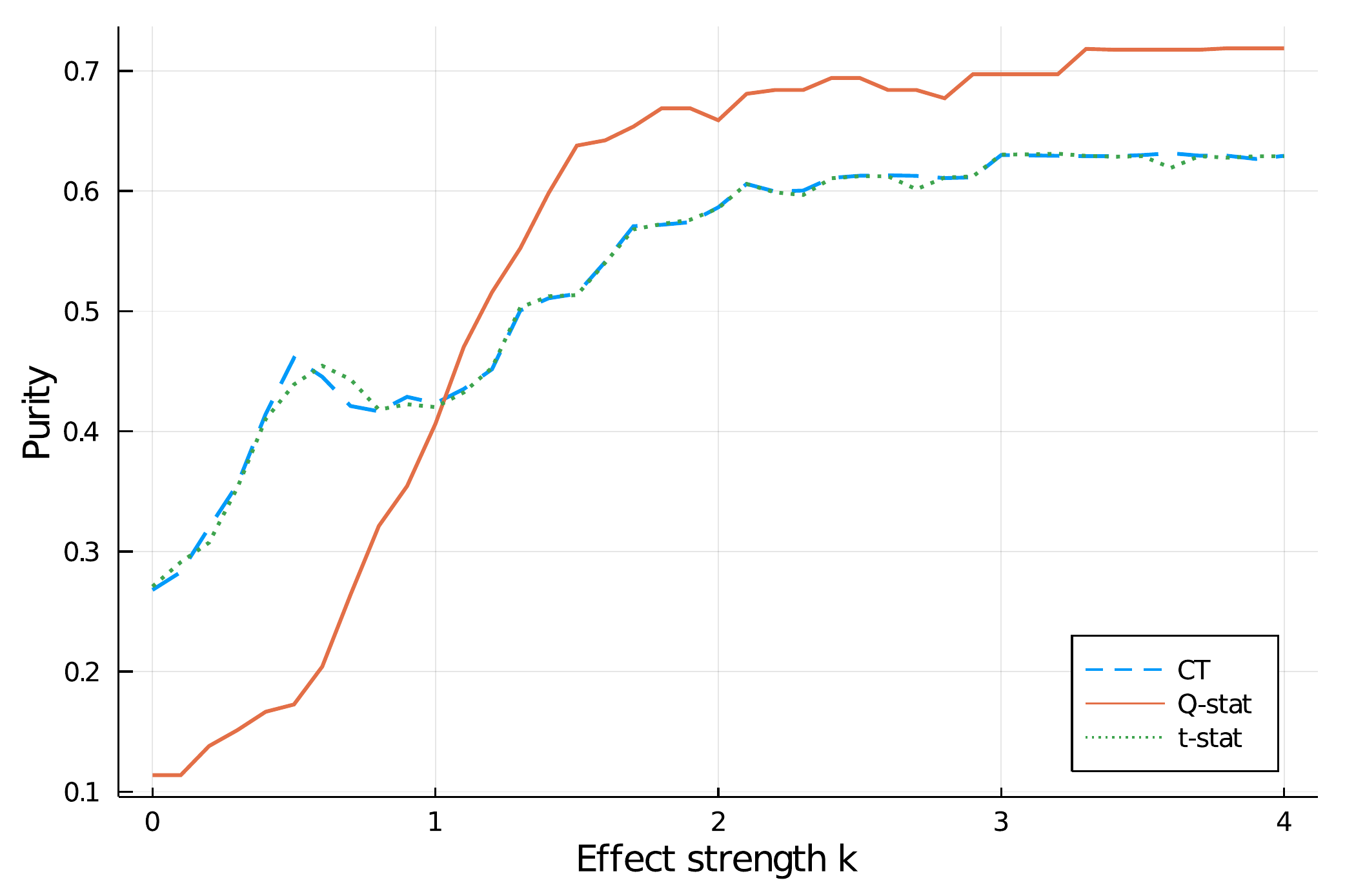}
    \end{subfigure}
    \caption{4-rule heterogeneous treatment effect}
    \label{fig:syn.4rule}
\end{subfigure}
\caption{Complexity (left panel) and purity (right panel) as the effect strength $k$ increases on synthetic examples with 2-rule (top) and 4-rule (bottom) heterogeneous treatment effect. Metrics are averaged over $20$ random datasets, containing $N=1,000$ observations each.}
\label{fig:syn}
\end{figure}

\paragraph{Results} Figure \ref{fig:syn.2rule} and \ref{fig:syn.4rule} report the results for the case where $\kappa$ comprises 2 and 4 rules respectively. From these results, we make the following observations: First, our procedure is a sound procedure to detect heterogeneous treatment effects. It returns partitions with low complexity and a purity comparable with what other methods from the literature achieve. As the strength of the treatment effect $k$ increases, purity of the leaves increases as well. Compared to the $t$-test and CT procedures, our Q-statistics criterion achieves better complexity, i.e., lower and less variable. This improvement is mostly due to fact that our hypothesis testing-based criterion softly penalizes the number of samples in each leaves, leading to relatively shallow trees. On the contrary, the two alternatives grow substantially deeper trees and require a pruning step which is a known source of instability (see Appendix \ref{sec:A.syn} for results obtained without pruning). Regarding purity, the performance of $t$-test and CT are remarkably similar. For low values of $k$, they dominate our approach by a substantial margin. This is not surprising since these methods were developed specifically for CATE estimation and leverage the fact that treatment effect can be expressed as a sample average. Our approach is more general, hence, less powerful in this setting, yet remarkably competitive. In particular, the edge of  $t$-test and CT shrinks as the effect strength or the number of rules increases. {Figure \ref{fig:syn.ortho} in appendix also suggests that Algorithm \ref{alg:rp} might be more robust to confounding variables.}
Together, our results suggest that our method is a viable and competitive alternative to the existing methods for HTE, in particular since it can accommodate for various measures of effect, such as odds ratios, unlike CT and $t$-test.

\section{Applications in Social Science and Medicine} \label{sec:appli}
In this section, we apply Algorithm \ref{alg:rp} to two data sets from social science and medicine.

\subsection{Social pressure and voting} \label{sec:appli.vote}
\citet{gerber2008social} conducted a large experiment to assess the effect of social pressure on voter turnout, by randomly sending mailings to 200,000 registered voters in the United States. We apply our HOR estimation procedure to mine for response heterogeneity, and compare our optimal treatment recommendations with prior analysis 
\citep{zhou2018offline}. 

\paragraph{Data description}
The dataset consists of $n=180,002$ voters from the state of Michigan. There are 10 voter characteristics: year of birth, sex, household size, city, and 6 voting indicator variables for the past three general and primary elections (2000, 2002, and 2004). \citet{gerber2008social} randomly {mailed a letter to the households before the 2006 primary election, with one of the four following messages (treatments):}
\begin{itemize}
    \item Civic Duty: ``Do your civic duty".
    \item Monitored : ``You are being studied". 
    \item Self History: The letter contains the household's past voting records. The letter also indicates that, once the election is over, a follow-up letter on whether the voter has voted will be sent to the household.
    \item Neighbors: The letter contains the voting records of the household and their neighbors, and also indicates that ``all your neighbors will be able to see your past voting records".
\end{itemize}
{A fifth group of households were not mailed anything (control group).}
The response to treatment is whether a voter has voted to the 2006 primary election (binary).

\paragraph{Results and insights}
We apply our recursive procedure for HOR estimation, in particular its multi-treatment extension developed in Section \ref{ssec:prescriptive}. We set $p_{\max}$ to $0.001$ and a maximal depth of 3. The resulting tree is displayed in Figure \ref{fig:tree.voter}. Consistent with the analysis of \citet{gerber2008social,zhou2018offline}, we find that voters display relatively homogeneous responses to mailings and that assigning ``Neighbors'' to everyone is the best policy. 
\newdimen\nodeDist
\nodeDist=40mm
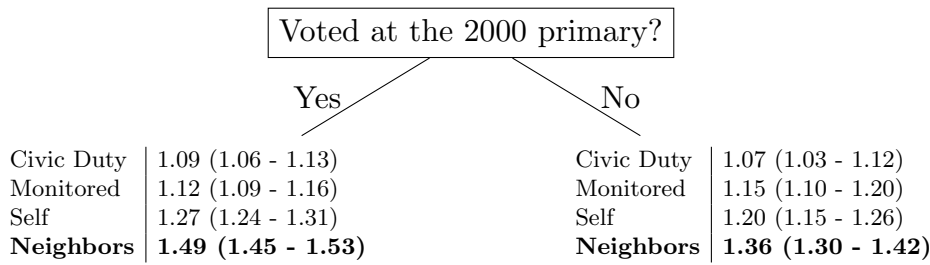
\begin{figure}
\centering
\begin{tikzpicture}[
    node/.style={%
      draw,
      rectangle,
    },
  ]
    \node [node] (A) {Voted at the 2000 primary?};
    \path (A) ++(-160:\nodeDist) node (B) [below]{
    \footnotesize
    \begin{tabular}{l|l}
         Civic Duty & 1.09 (1.06 - 1.13) \\
         Monitored & 1.12 (1.09 - 1.16) \\
         Self & 1.27 (1.24 - 1.31) \\ 
         \bf Neighbors & \bf 1.49 (1.45 - 1.53)
    \end{tabular}
     };
    \path (A) ++(-20:\nodeDist) node (C) [below]{   \footnotesize
    \begin{tabular}{l|l}
         Civic Duty & 1.07 (1.03 - 1.12) \\
         Monitored & 1.15 (1.10 - 1.20) \\
         Self & 1.20 (1.15 - 1.26) \\ 
         \bf Neighbors & \bf 1.36 (1.30 - 1.42)
    \end{tabular}
     };
    \draw (A) -- (B) node [left=5pt,pos=0.5] {Yes}(A);
    \draw (A) -- (C) node [right=5pt,pos=0.5] {No}(A);
\end{tikzpicture}
\caption{Best partition of depth at most 3 obtained for measuring heterogeneity in response to social pressure for the voting dataset.}
\label{fig:tree.voter}
\end{figure}

\subsection{Application on a medicine example} \label{sec:appli.cancer}
We also illustrate our approach on medical data from a large clinical trial conducted in the United States starting in 2006. 

\paragraph{Data description} The Prostate, Lung, Colorectal and Ovarian (PLCO) Cancer Screening Trial is a large randomized trial designed and sponsored by the National Cancer Institute to determine the effects of screening on cancer-related mortality. We restrict our attention to colorectal cancer and conduct two analyses, namely evaluate the impact of treatment (i.e., screening) on the incidence of colorectal cancer and the corresponding cancer stage. The trial comprises 154,887 individuals, $50.5\%$ female, aged between 42 and 78 years old at randomization. Out of them, 2,299 contracted and/or died of colorectal cancer. The covariates we use for heterogeneity detection correspond to data available at the beginning of the trial, namely age, gender, prior history of colorectal or any cancer, and eligibility to colorectal analysis based on the Baseline Questionnaire. 

\paragraph{Results and insights} We first consider the entirety of the cohort and evaluate the impact of screening on overall cancer incidence. At a population level, we do observe that intervention leads to lower incidence of colorectal cancer, with odds ratio 0.7984 (95\% CI: 0.7387 - 0.8630), with a stronger effect on men than women (see Figure \ref{fig:tree.plco.incidence}). For the 2,299 patients who contracted a colorectal cancer, we study whether the screening impacted the cancer stage at detection. Indeed, an earlier stage cancer is associated with higher survival \citep{bannister2016cancer}. 
Accordingly, we consider a stage I or II cancer as a ``positive'' outcome. Screening does have an overall positive impact on cancer stage (OR: 0.98 - 1.36). Yet, it is important to notice that we observe no significant effect on  patients aged less than 60 years old (Figure \ref{fig:tree.plco.stage}). From a policy perspective, this analysis suggests that enforcing mandatory screening before 60 years old would be inefficient in improving cancer screening and that screening efforts should focus on $60+$ year-old individuals.   
\newdimen\nodeDist
\nodeDist=25mm
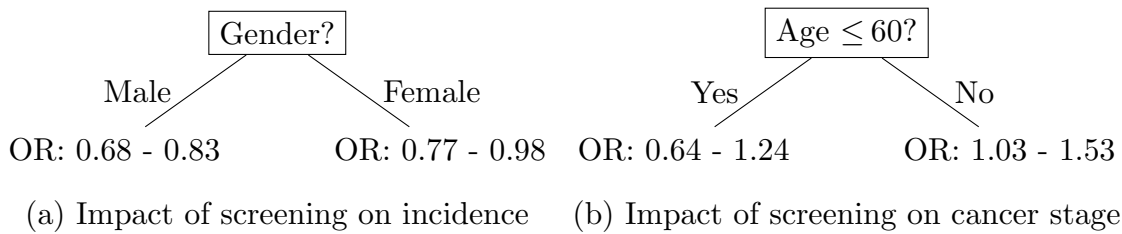
\begin{figure}
\centering
\begin{subfigure}[b]{.45\textwidth}
\centering
\begin{tikzpicture}[
    node/.style={%
      draw,
      rectangle,
    },
  ]
    \node [node] (A) {Gender?};
    \path (A) ++(-150:\nodeDist) node (B) [below]{OR: 0.68 - 0.83};
    \path (A) ++(-30:\nodeDist) node (C) [below]{OR: 0.77 - 0.98};
    \draw (A) -- (B) node [left=5pt,pos=0.5] {Male}(A);
    \draw (A) -- (C) node [right=5pt,pos=0.5] {Female}(A);
\end{tikzpicture}
\caption{Impact of screening on incidence}
\label{fig:tree.plco.incidence}
\end{subfigure}
\begin{subfigure}[b]{.45\textwidth}
\centering
\begin{tikzpicture}[
    node/.style={%
      draw,
      rectangle,
    },
  ]
    \node [node] (A) {Age $\leq 60$?};
    \path (A) ++(-150:\nodeDist) node (B) [below]{OR: 0.64 - 1.24};
    \path (A) ++(-30:\nodeDist) node (C) [below]{OR: 1.03 - 1.53};
    \draw (A) -- (B) node [left=5pt,pos=0.5] {Yes}(A);
    \draw (A) -- (C) node [right=5pt,pos=0.5] {No}(A);
\end{tikzpicture}
\caption{Impact of screening on cancer stage}
\label{fig:tree.plco.stage}
\end{subfigure}
\caption{Outcome of the heterogenous odds ratio estimation procedure on the PLCO data set. Algorithm \ref{alg:rp} is set with a maximal depth of 3 and a $p_{max}$ value of $0.1$.}
\end{figure}

\section{Application on a Revenue Management Example} \label{sec:jd}
We apply the proposed R-HOR method to data from the e-tailer JD.com \citep{shen2019jd} in order to study the effect of discounts on the purchase probability.

\subsection{Data description and processing}
We aggregated data from different parts of the dataset, which we describe in Appendix \ref{ssec:A.jdcom.process}. We refer to \citet{shen2019jd} for a more detailed presentation of the data and summary statistics. 
The final dataset consists of $566,004$ customer-product interactions, described with 9 variables concerning the user (number of past purchase, time since first purchase, ``Plus'' membership indicator, age, gender, education, marital status, purchase power category, city level), 4 variables concerning the product (product type, original price, attribute 1 and 2 missingness indicator), and 2 variables regarding the interaction (day of the week, hour of the day). For each interaction, we know whether the customer eventually purchased the product. If so, we know whether they were offered a discounted price.

\subsection{Predicting discount assignment}
Out of all $566,004$ customer-product interactions, $42,184$ of them constitute a transaction. $61\%$ of these transactions were offered a discount. We use this data to train a machine learning model to predict the probability of being treated, i.e., discounted, given the $15$ observed covariates, $\mathbb{P}\left(T=1 | \bm{X} = \bm{x}, Y=1 \right)$. We randomly split the data into a training ($70\%$) and a test set ($30\%$). We calibrate all models on the training data, using $3$-fold cross validation for tuning hyper-parameters. We compare a single decision tree \citep{breiman1984classification}, random forest \citep{breiman2001random} and nearest neighbor classification \citep{cover1967nearest} and report their out-of-sample $AUC$ 
in Table \ref{tab:acc}. We acknowledge that this comparison is purely illustrative and not exhaustive.
\begin{table}
\footnotesize
\caption{Out-of-sample performance of machine learning algorithms for predicting discount assignment} \label{tab:acc}
\centering
\begin{tabular}{llc}
Model & Hyper-parameters & Out-of-sample $AUC$ \\ 
\toprule
Decision tree & maximal depth, criterion & 0.7163 \\ 
Random forest & maximal depth, criterion, number of trees & 0.7295 \\ 
Nearest neighbors & neighborhood size, norm & 0.6914 \\ 
\bottomrule
\end{tabular}
\end{table}

As summarized in Table \ref{tab:acc}, all machine learning models can predict discount assignment with an $AUC$ in the low $0.70$s.  
Out of the three, random forest is the most accurate model and the one we use in the remaining analysis. Among others, machine learning models such as these ones can be used to identify the drivers of discounts in the platform's current promotion targeting strategy (see, e.g., Table \ref{tab:drivers} in the online appendix).

\subsection{Imputing discount assignments}
We now apply the previously-trained random forest model to all observations in our data set corresponding to unbought items, thus obtaining a score $\hat{t}_i \in [0,1]$ for each item. 

Under the MAR assumption, $\hat{t}_i$ estimates the probability $\mathbb{P}( T = 1 | \bm{X} = \bm{x}_i )$. Accordingly, for unbought items, we can either consider a single data set where unbought items are softly assigned to treatment, i.e., $(\bm{x}_i, \hat{t}_i, y_i=0)$, or consider $N_s$ potential treatment assignments sampled according to $\hat{\bm{t}}$, i.e., $(\bm{x}_i, {t}_i^{(j)}, y_i=0)$,  $j=1,\dots,N_s$, with ${t}_i^{(j)} \sim \mathrm{Bernoulli}(\hat{t_i})$. We will refer to these options as ``MAR'' and ``MAR sampled'' respectively. We can also impute the missing treatment assignments using chained equations \citep{buuren2010mice}, a widely recognized technique for missing data imputation, implemented in the \verb|R| package \verb|mice|. We perform 5 iterations of the algorithm to generate 5 different imputed data set (default configuration) and refer to this method as ``MICE''. Finally, we compare these approaches with the methodology presented in Section \ref{ssec:rob.sample}, where we consider $N_s$ potential treatment assignments $\hat{\bm{t}}^{(j)} \in \mathcal{T}$. We fix $N_{s}=160$, which, according to Proposition \ref{prop:sampling}, will lead to a violation probability $\epsilon$ and a confidence parameter $\beta$ of 5\%.

We compare the four imputation methods in Table \ref{tab:impute}. First, we observe that ``MAR'' and ``MAR-sampled'' lead to similar results, which is justified by the fact that odds ratio can account for fractional treatment assignment. Second, methods relying on the MAR assumption, namely ``MAR'', ``MAR-sampled'' and ``MICE'', provide similar results: they replicate the treatment assignment mechanism observed on the purchased items. As a result, the proportion of discounted items is similar to the one observed on the bought items ($61.1\%$) and the overall odds ratio is close to 1. In addition, multiple imputation methods (``MAR-sampled'' and ``MICE'') produce data sets which tend to be very similar. On the contrary, our less restrictive set of assumptions lead to a greater diversity of scenarios, which is both a blessing and a curse. On the bright side, it allows for cases where discounts can have a (strong) effect.  However, scenarios will be generated adversarially within the HOR estimation algorithm so a larger set of treatment assignments also provides the adversarial with more power to revoke our findings.  
\begin{table}
\footnotesize
\caption{Comparison of the different imputation methods in terms of fraction of discounted items and overall odds ratio. For methods producing multiple imputed data sets, we report average values (and standard deviation)}
\label{tab:impute}
\centering
\begin{tabular}{lcccc}
 & MAR & MAR - sampled & MICE & $\mathcal{T}$ \\ 
\toprule
Fraction of discounts & $62.9\%$ &  $62.9\% \ (0.1\%)$ & $63.7\% \ (0.4\%)$ & $34.9\% \ (12.6\%)$ \\
Odds ratio & $0.929$  &  $0.929 \ (0.003)$   &  $0.895 \ (0.015)$  &  $4.072 \ (3.994)$ \\
\bottomrule
\end{tabular}
\end{table}

\subsection{Heterogeneous odds ratio estimation}
We first consider the vector of treatment assignments $\bm{t}$ leading to the worst odds ratio, i.e., the closest to $1$. By doing so, we estimate that offering a discount slightly increases the likelihood of purchase, with an odds ratio of 1.12 (95\% CI: 1.10-1.14)--see the contingency table, Table \ref{tab:cont.jd}, in the online appendix. 

This treatment assignment serves as the starting point for our robust heterogeneous assignment. Figure \ref{fig:jd.tree1} displays the best tree of depth 1 found with this treatment assignment. In particular, we observe that discounts are more effective on cheap products, i.e., product that cost less than 36 (currency unknown). Following the robust estimation procedure (Algorithm \ref{alg:robrp}), this split has to be validated on a range of possible treatment assignment values by solving Problem \eqref{eqn:rob.tradeoff}. Figure \ref{fig:jd.budget} displays the worst $p$-value for Cochran'$Q$ test applied on this split for different  budgets of uncertainty $\Gamma$. The higher the $\Gamma$, the larger the range of potential treatment assignments $\mathcal{T} \cap \{\bm{t} : \| \bm{t} - \bm{t}_0 \|_1 \leq n \Gamma \}$, the higher the $p$-value can be. In the remaining of the analysis, we fix $p_{max}=0.01$ and $\Gamma=0.15$.
\newdimen\nodeDist
\nodeDist=25mm
\begin{figure}
\centering
\begin{subfigure}[b]{.45\textwidth}
\centering
\begin{tikzpicture}[
    node/.style={%
      draw,
      rectangle,
    },
  ]
    \node [node] (A) {Product type?};
    \path (A) ++(-150:\nodeDist) node (B) [below]{OR: 2.78 - 3.92};
    \path (A) ++(-30:\nodeDist) node (C) [below]{OR: 1.09 - 1.14};
    \draw (A) -- (B) node [left=5pt,pos=0.5] {1P}(A);
    \draw (A) -- (C) node [right=5pt,pos=0.5] {3P}(A);
\end{tikzpicture}
\caption{Best tree found of depth 1}
\label{fig:jd.tree1}
\end{subfigure}
\begin{subfigure}[b]{.45\textwidth}
\centering
\includegraphics[width=\textwidth]{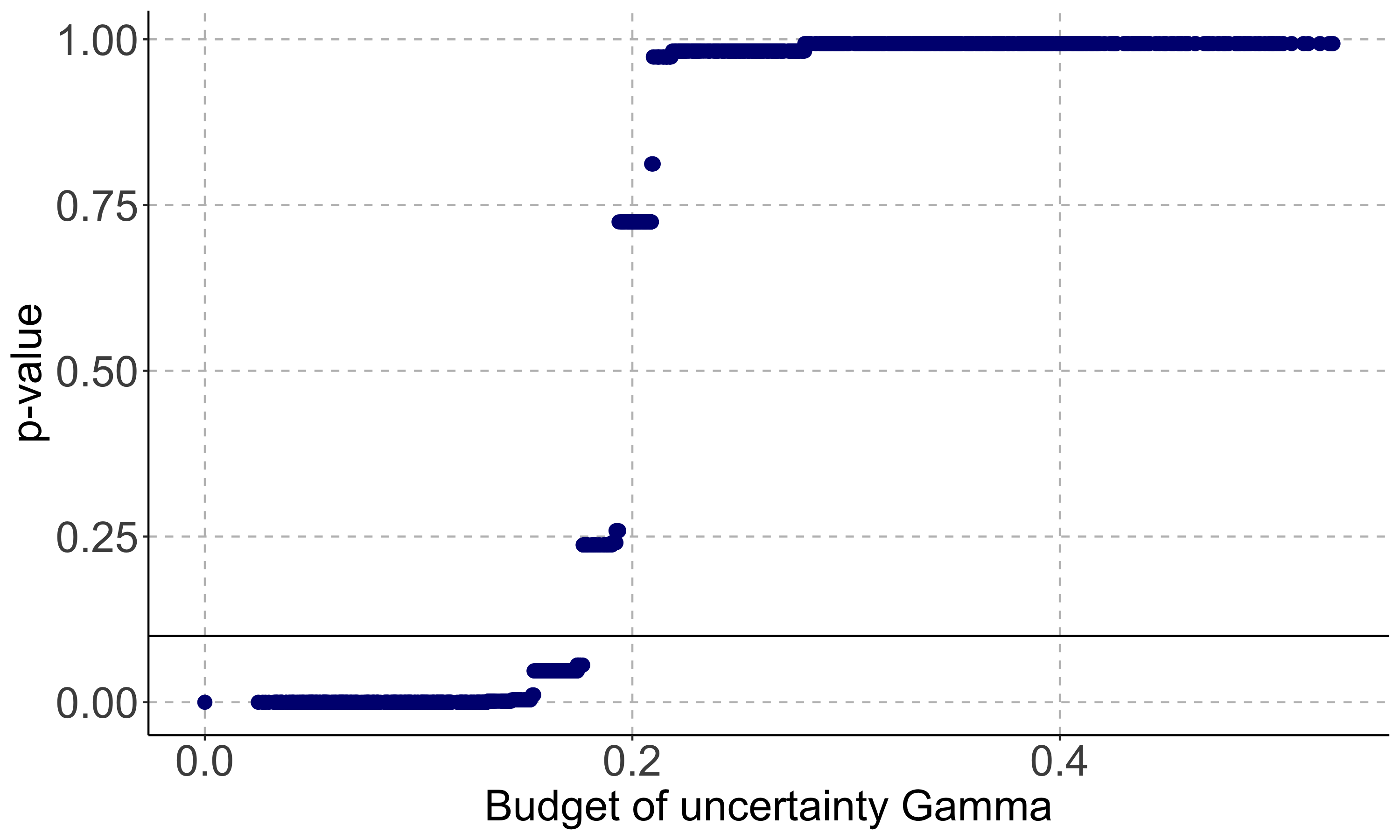}
\caption{Impact of $\Gamma$ on the resulting $p$-value}
\label{fig:jd.budget}
\end{subfigure}
\caption{Outcome of the heterogenous odds ratio estimation procedure on the JD.com data set. Algorithm \ref{alg:rp} is set with a maximal depth of 1 and a $p_{max}$ value of $0.01$.}
\end{figure}

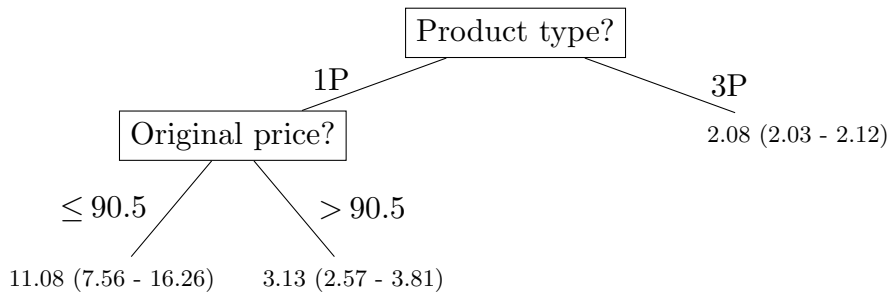
\begin{figure}
\centering
\begin{tikzpicture}[
    node/.style={%
      draw,
      rectangle,
    },
  ]
    \node [node] (A) {Product type?};
    \path (A) ++(-160:40mm) node[rectangle,draw] (B) {Original price?};
    \draw (A) -- (B) node [above=2pt,left=5pt,pos=0.5] {1P};

    \path (B) ++(-130:25mm) node (BL) {\footnotesize 11.08 (7.56 - 16.26)};
    \draw (B) -- (BL) node [left=5pt,pos=0.5] {$\leq 90.5$};
    
    \path (B) ++(-50:25mm) node (BR) {\footnotesize 3.13 (2.57 - 3.81)};
    \draw (B) -- (BR) node [right=5pt,pos=0.5] {$> 90.5$};

    \path (A) ++(-20:40mm) node (C) {\footnotesize 2.08 (2.03 - 2.12)};
    \draw (A) -- (C) node [right=15pt,pos=0.5] {3P};
\end{tikzpicture}
\caption{Best partition of depth at most 3 obtained for measuring heterogeneity in response to discount on JD.com data.}
\label{fig:jd.tree}
\end{figure}

Figure \ref{fig:jd.tree} displays the best robust partition obtained of depth at most 3. From this partition, we can make the following observations: First, discount has a overall positive impact on purchase probability. Second, product-related variables lead to more heterogeneity in discount effectiveness than customer-related information, suggesting that customers can be relatively homogeneous in their response to discounts. However, their behavior is strongly dependent on the product being discounted. Finally, Two categories of products benefit the most from promotions: cheap products and products from a ``1P'' supplier. 
\newdimen\nodeDist
\nodeDist=40mm

\section{Conclusion} 
Motivated by transaction-level data from a major e-tailer, we propose a recursive partitioning procedure for estimating heterogeneous odds ratio,  
a widely used measure in medicine, economics, and social sciences. 
Our procedure can be applied to any definition of the treatment effect for which an asymptotically normal estimator exists. In addition, we develop an adversarial imputation procedure to allow for robust estimation in presence of partially observed treatment assignment, a central problem in data-driven revenue management. We believe that the integration of problem-specific imputation methods directly into statistical estimation, and their theoretical analysis, constitutes a promising future direction for analytics in operations management.
  
\ACKNOWLEDGMENT{%
The authors would like to thank JD.com and the M\&SOM 2020 Data Driven Research Challenge committee for offering access to transaction-level data and administering this research challenge, which motivated this work. The authors also thank the National Cancer Institute for access to NCI's data collected by the Prostate, Lung, Colorectal and Ovarian (PLCO) Cancer Screening Trial analyzed in Section \ref{sec:appli.cancer}. The statements contained herein are solely those of the authors and do not represent or imply concurrence or endorsement by NCI. Thanks to Ali Aouad, Colin Fogarty, and Nicos Savva for providing early feedback on the work and suggesting relevant references, and to two anonymous referees for their constructive comments.
}


\bibliographystyle{informs2014} 
\bibliography{biblio.bib} 

\newpage 
\ECSwitch
\renewcommand{\thesection}{A.\arabic{section}}%
\renewcommand{\thefigure}{A.\arabic{figure}}%
\setcounter{table}{0}%
\setcounter{figure}{0}%
\setcounter{algocf}{0}
\renewcommand\thetable{A.\arabic{table}}%
\setcounter{page}{1}\def\thepage{a\arabic{page}}%
\renewcommand{\thealgocf}{A.\arabic{algocf}}


\ECHead{Online Appendix \\ Pauphilet, {\it Robust and Heterogenous Odds Ratio: Estimating Price Sensitivity for Unbought Items}}

\section{Recursive procedure for Heterogeneous Odds Ratio estimation}
We provide in this section the pseudo-code of the recursive partitioning procedure for heterogeneous odds ratio estimation described in Section \ref{ssec:hor.alg}.

The algorithm is initialized with the entire space as the initial partition, $\Pi = \{\mathbb{R}^p \}$. 
At each step, the best split is found by exploring the $p$ features and all the potential values for $t$. For continuous variables, this search can be accelerated by searching threshold values among a predefined list of quantiles as in \citet{aouad2019market}.  
Finally, depth of the leaf or number of training samples in the leaf (total or by treatment regime) can be used as extra criteria to declare a leaf unsplittable, although they are already indirectly captured in the $Q$-statistics threshold. To account for multiple hypothesis testing, the threshold value $p_{\max}$ could be reduced throughout the algorithm.

\begin{algorithm}[h]
\footnotesize
\SetAlgoLined
\KwIn{$\mathcal{D}_n = \{(\bm{x}_i, t_i, y_i),\, i=1,\dots,n\}$}
\KwResult{Partition $\Pi$.}
 initialization: $\Pi \leftarrow \{\mathbb{R}^p \}$\;
 \For{$\ell \in \Pi$}{
 $(j^\star, t^\star) \leftarrow \argmax_{(j,t)} \: Q(\{\ell_-,\ell_+\} ,\mathcal{D}_n)$ \;
  \If{$1 - \chi_1^2(Q(\{\ell^\star_-,\ell^\star_+ \},\mathcal{D}_n)) \leq p_{\max}$  \texttt{and}  other criterion}
  {split $\ell$: $\Pi \leftarrow (\Pi \backslash \{\ell \}) \cup \{\ell_-,\ell_+ \} $}
 }
\caption{Recursive partitioning algorithm for heterogeneous odds ratio estimation} \label{alg:rp}
\end{algorithm}

\section{Discussion: From trees to forest} \label{ssec:t.rf}
Due to the discrete nature of the splits, decision trees can be notoriously unstable to perturbations in the data. To improve stability, \citet{breiman2001random} originally proposed to introduce randomization in the training process of each tree, train a collection of random trees, and aggregate them in a random forest. \citet{wager2018estimation} later extended this idea for HTE estimation. 
{\citet{wager2015adaptive} leveraged their adaptive concentration bounds over trees to derive similar results for random forests. While a thorough analysis of heterogeneous odds ratio estimation via random forest is out of the scope of the present paper, we discuss in this section the different alternatives available to extend our methodology to random forests.}

{\bf Source of randomness:} Following \citet{breiman2001random}, randomness can be introduced by calibrating each tree on a bootstrap sample of the training data $\mathcal{D}_n$. Randomness can also be introduced in the split selection process, for instance by searching the best split among a random subset of covariates (typically, of size $\sqrt{p}$), {or by imposing that at each iteration  all variables can be selected with probability at least $\pi / p$ for some $\pi \in (0,1]$. The later assumption was proposed by \citet{meinshausen2006quantile} to ensure that the size of the leaves shrinks as $n \to \infty$ and guarantee consistency.} Note that stability of decision trees can also be improved by introducing randomness in the nature of the splits directly instead of the training process \citep[see][]{yuan1995induction,last2002improving,benard2019sirus,perakis2019xstrees}.

{\bf Aggregation:} Typically, the output of a random forest is obtained by averaging. However, given the non-collapsibility of odds ratios, the outputs of each tree in a forest cannot be simply averaged. Consider an observation $\bm{x}$ and a random forest consisting of $n_{trees}$ trees. Each tree $\tau$ places $\bm{x}$ in a leaf $\ell(\bm{x},\tau)$. Let us denote $\pi_{t,y}^{(\tau)}(\bm{x})$ the value of $\hat{\pi}^n_{t,y}(\ell(\bm{x},\tau))$, and $N^{(\tau)}(\bm{x})$ the number of training samples in that leaf.
{By generalizing Lemma \ref{lemma:wager} to random forests as in \citet{wager2015adaptive}, we can show that $\tfrac{1}{n_{tree}} \sum_\tau \pi_{t,y}^{(\tau)}(\bm{x})$ provides consistent estimates of $\pi_{t,y}(\bm{x})$. Hence, under the assumptions of Theorem \ref{thm:or},
\begin{align*}
    \dfrac{\left( \sum_\tau \pi^{(\tau)}_{1,1}(\bm{x}) \right) \left( \sum_\tau \pi^{(\tau)}_{0,0}(\bm{x}) \right)}{\left( \sum_\tau \pi^{(\tau)}_{1,0}(\bm{x}) \right) \left( \sum_\tau \pi^{(\tau)}_{0,1}(\bm{x}) \right)}
\end{align*}
is a consistent estimate of the odds ratio at $\bm{x}$. Note that this aggregation rule differs from} the Mantel-Haenszel estimator \citep{mantel1959statistical}:
\begin{align*}
    \dfrac{\sum_{\tau} \pi^{(\tau)}_{1,1}(\bm{x}) \pi^{(\tau)}_{0,0}(\bm{x}) N^{(\tau)}(\bm{x}) }{\sum_{\tau} \pi^{(\tau)}_{1,0}(\bm{x}) \pi^{(\tau)}_{0,1}(\bm{x}) N^{(\tau)}(\bm{x})  }.
\end{align*}
We refer to \citet{breslow1981odds,breslow1982variance} for a discussion on the estimation of its variance. 
However, the Mantel-Haenszel estimator is only valid if leaves of different trees can be interpreted as independent strata, which is not the case if they are obtained via bootstrapped sample of the same training data. Finally, under the assumption that the response follows a logistic model,
\begin{align*}
    \log \: \dfrac{\mathbb{P}(Y = 1|\bm{X}=x, T=t)}{\mathbb{P}(Y = 0|\bm{X}=x, T=t)} = r(\bm{x}) t + \bm{\nu}(\bm{x})^\top \bm{x},
\end{align*}
the odds ratio corresponds to the coefficient associated with the treatment variable, $r(\bm{x})$. \citet{athey2019generalized} extended the random forest methodology to estimate such non-parametric generalized linear model (here, the functions $r$ and $\bm{\nu}$ are non parametrized): For each $\bm{x}$, $r(\bm{x})$ can be estimated by solving a weighted maximum likelihood estimation problem, where each training sample is weighted according to membership in the leaves of the random forest $\bm{x}$ belongs to. \citet{oprescu2019orthogonal} later proposed an orthogonalized version of the procedure that is robust to the estimation error in the nuisance component $\nu(\bm{x})$. Besides positing a logistic model, the main limitation of these approaches, however, is that the odds ratio is not explicitly provided by the random forest but rather as the solution of an optimization problem that depends on the random forest output.

\section{Uniform sampling procedure for treatment assignments} \label{sec:A.sampling}
Given scores $\hat{\bm{t}} \in [0,1]^n$, we randomly sample from $\mathcal{T}$ following Algorithm \ref{alg:sampling}.
\begin{algorithm}[h]
\footnotesize
\SetAlgoLined
\KwIn{$ \hat{\bm{t}} \in [0,1]^n$}
\KwResult{random sample $\bm{t} \in \mathcal{T}$.}
 initialization: $t_i \leftarrow \verb|NA|$, $\sigma \leftarrow \mathrm{sortperm}(\hat{\bm{t}})$\;
 \While{$\{i : t_i = \mathtt{NA} \} \neq \emptyset$}{
 $i \leftarrow \mathrm{random}(\{i : t_i = \mathtt{NA} \})$ \; 
 $\mathcal{I}_- \leftarrow \{j : \sigma(j) < \sigma(i),\ t_j \neq \mathtt{NA} \}$, %
 $\mathcal{I}_+ \leftarrow \{j : \sigma(j) > \sigma(i),\ t_j \neq \mathtt{NA} \}$ 
 
 $LB \leftarrow \max( t_j, j \in \mathcal{I}_-)$, $UB \leftarrow \min( t_j, j \in \mathcal{I}_+)$ \tcp{Note: if $\mathcal{I}_- $ (resp. $\mathcal{I}_+$) empty, 0 (resp. 1).} 
 $UB \leftarrow \min(UB, \hat{t}_i)$;
 $t_i \leftarrow \mathcal{U}([LB,UB])$\;
 }
\caption{Uniform sampling procedure} \label{alg:sampling}
\end{algorithm}

\FloatBarrier
\section{Additional numerical experiments on synthetic data} \label{sec:A.syn}
\begin{figure}
\begin{subfigure}{\linewidth}
    \centering
    \begin{subfigure}{.45\linewidth}
    \includegraphics[width=\linewidth]{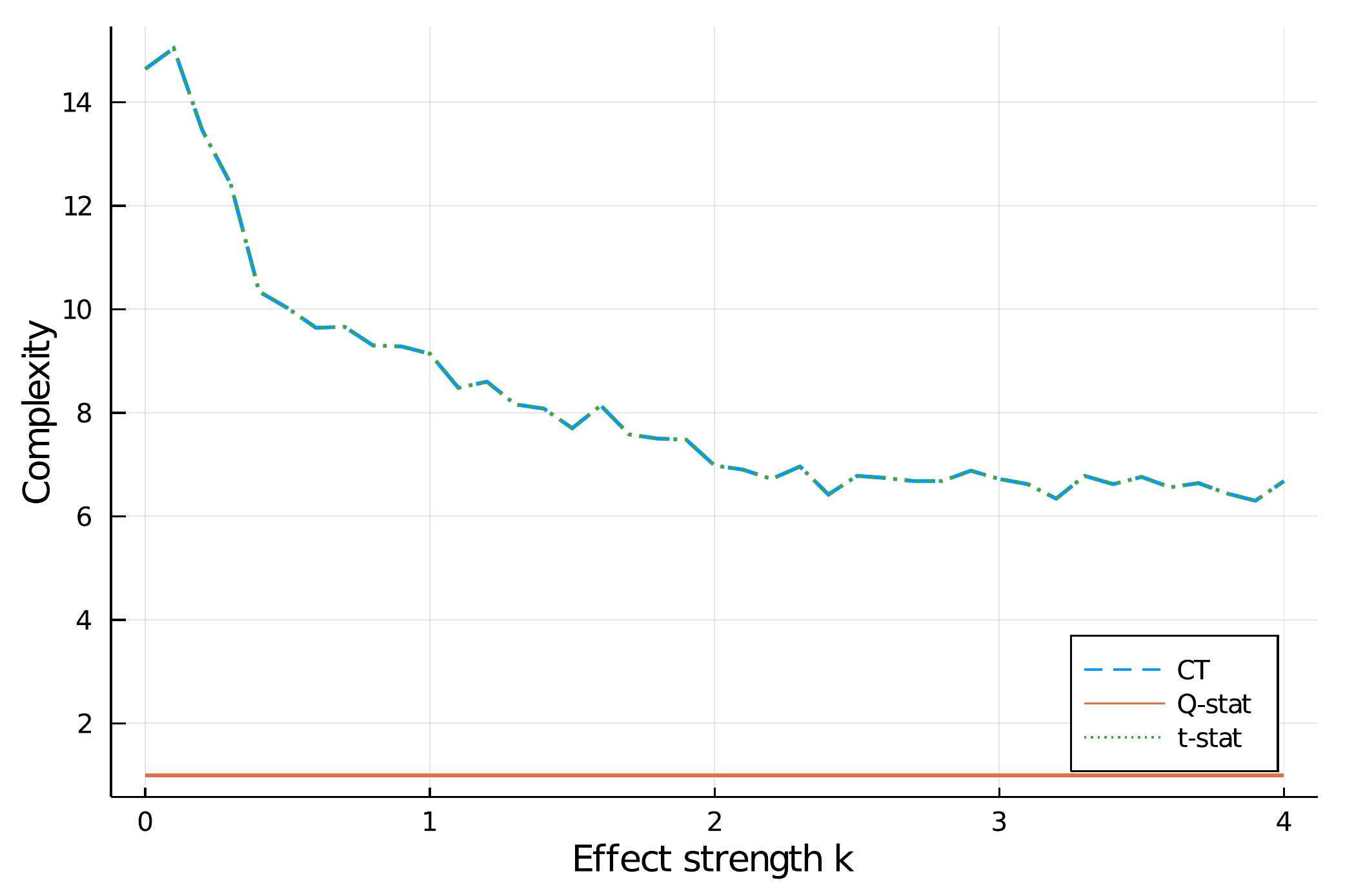}
    \end{subfigure} %
    \begin{subfigure}{.45\linewidth}
    \includegraphics[width=\linewidth]{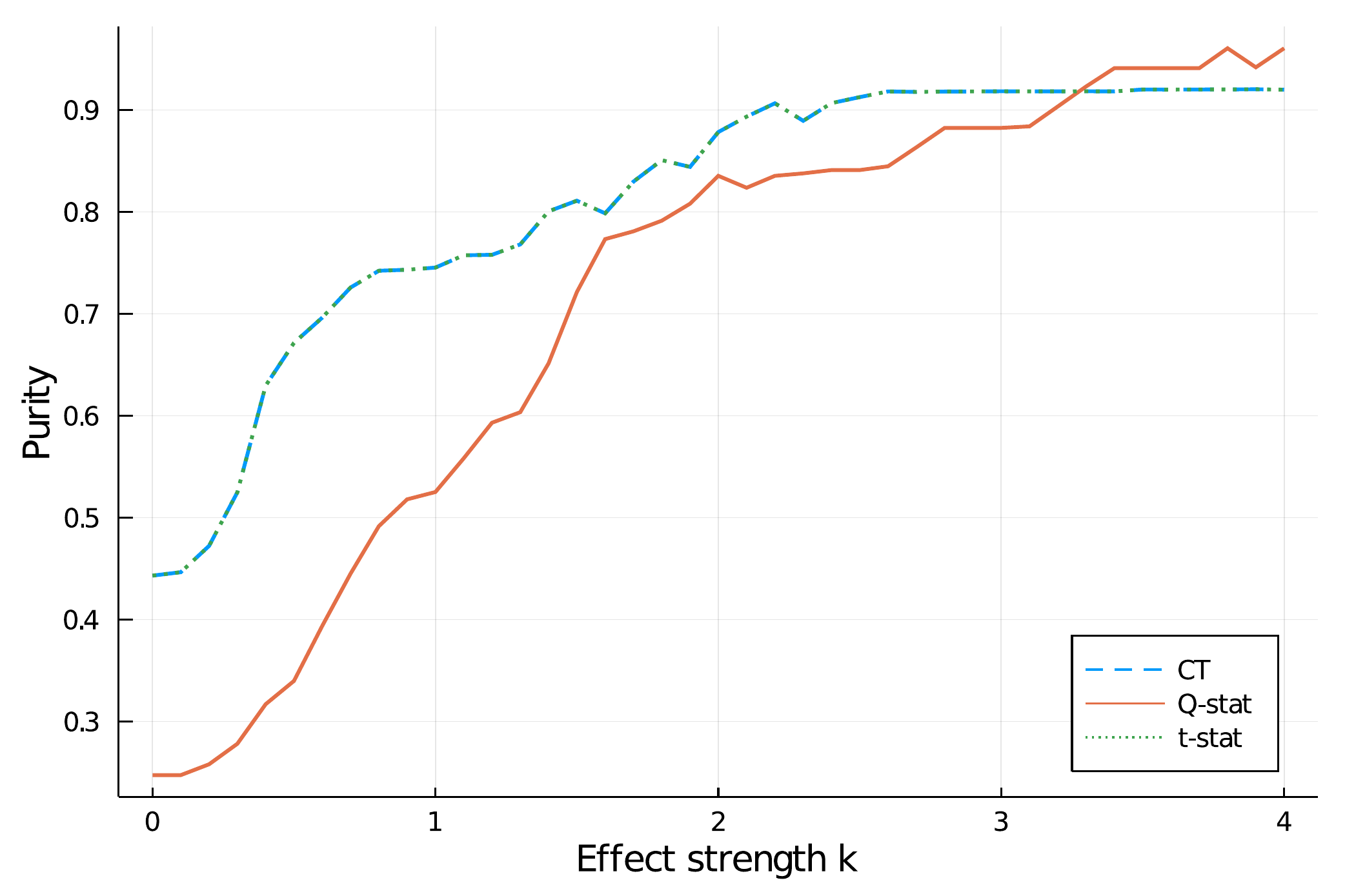}
    \end{subfigure}
    \caption{2-rule heterogeneous treatment effect.}
    \label{fig:syn.2rule.nopruning}
\end{subfigure}
\begin{subfigure}{\linewidth}
    \centering
    \begin{subfigure}{.45\linewidth}
    \includegraphics[width=\linewidth]{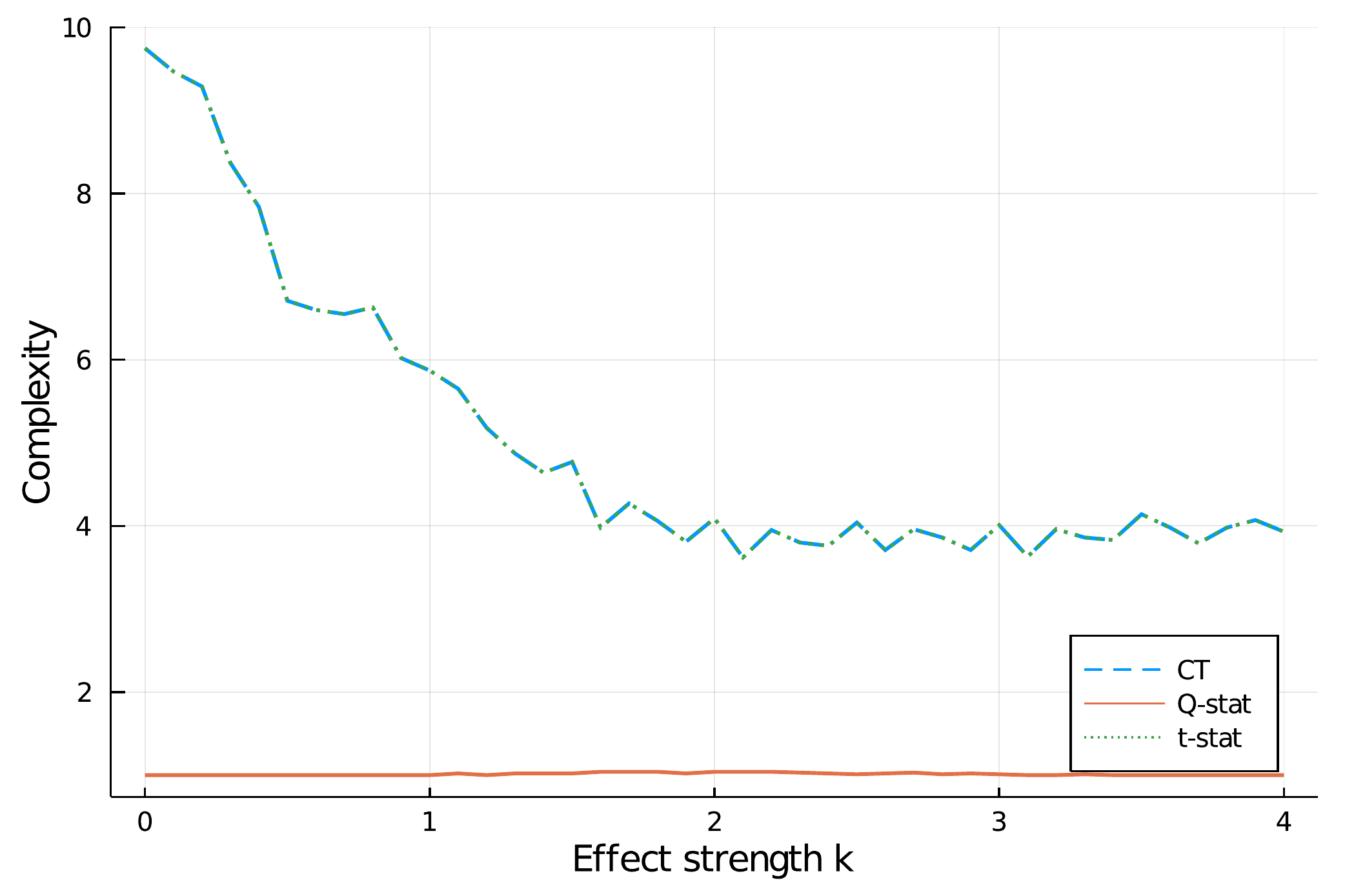}
    \end{subfigure} %
    \begin{subfigure}{.45\linewidth}
    \includegraphics[width=\linewidth]{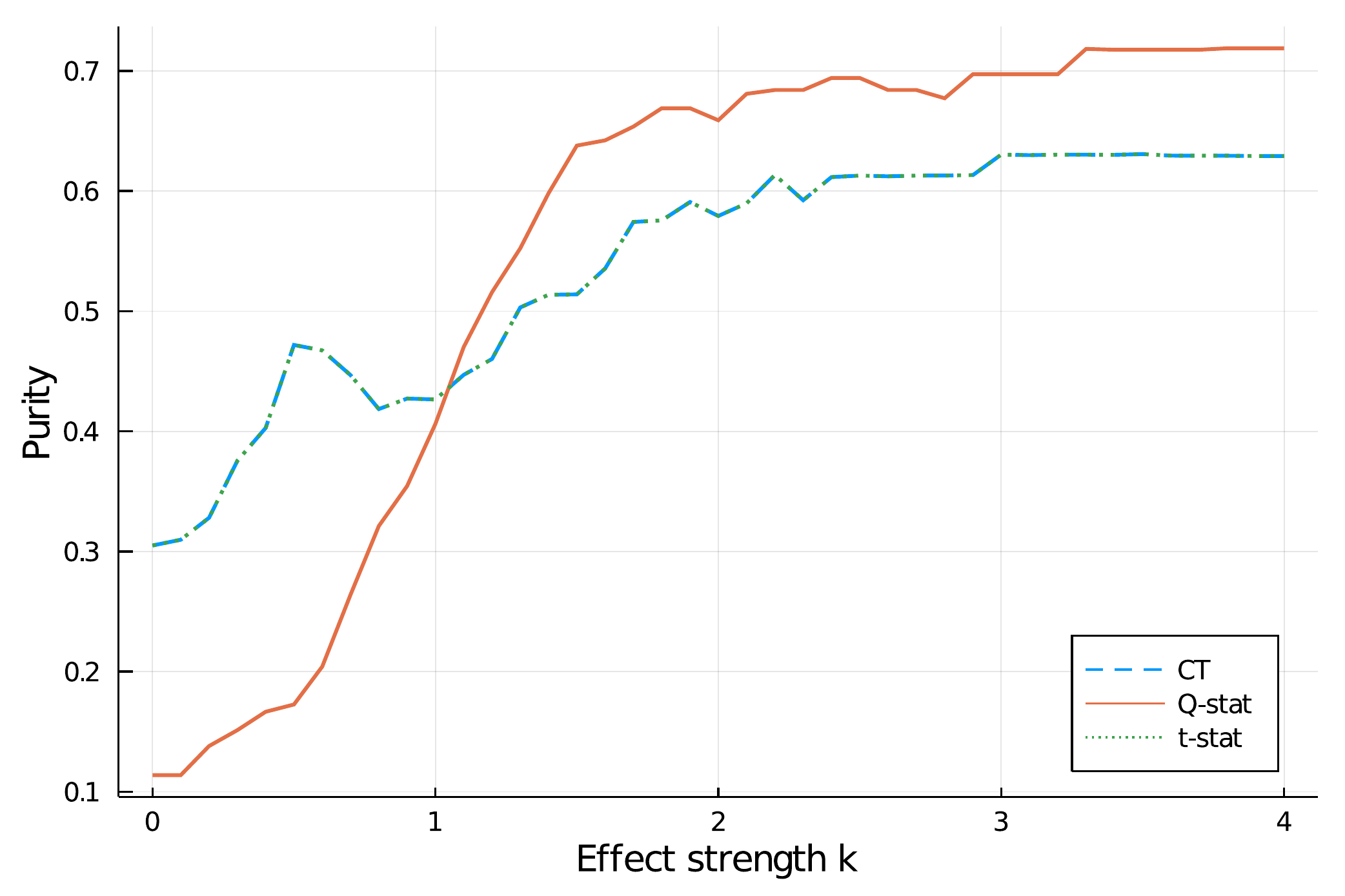}
    \end{subfigure}
    \caption{4-rule heterogeneous treatment effect.}
    \label{fig:syn.4rule.nopruning}
\end{subfigure}
    \caption{Replication of Figure \ref{fig:syn} when trees obtained by ``$t$-test'' and ``CT'' are not pruned. The left and right panel respectively display the complexity and purity as the effect strength $k$ increases, on synthetic examples with 2-rule (top) 4-rule (bottom) heterogeneous treatment effect.  Metrics are averaged over $20$ random datasets, containing $N=1,000$ observations each.}
    \label{fig:syn.nopruning}
\end{figure}

Figures \ref{fig:syn.2rule.nopruning} and \ref{fig:syn.4rule.nopruning} replicate Figures \ref{fig:syn.2rule} and \ref{fig:syn.4rule} from Section \ref{sec:syn}, except that trees obtained by ``$t$-test'' and ``CT'' are not pruned. These results clearly demonstrate the importance of pruning in reducing the size of the trees, hence achieving lower complexity, without compromise on their quality (purity almost unchanged). Yet, it is also clear from these pictures that pruning introduces instability. 

\begin{figure}
    \centering
    \begin{subfigure}{.45\linewidth}
    \includegraphics[width=\linewidth]{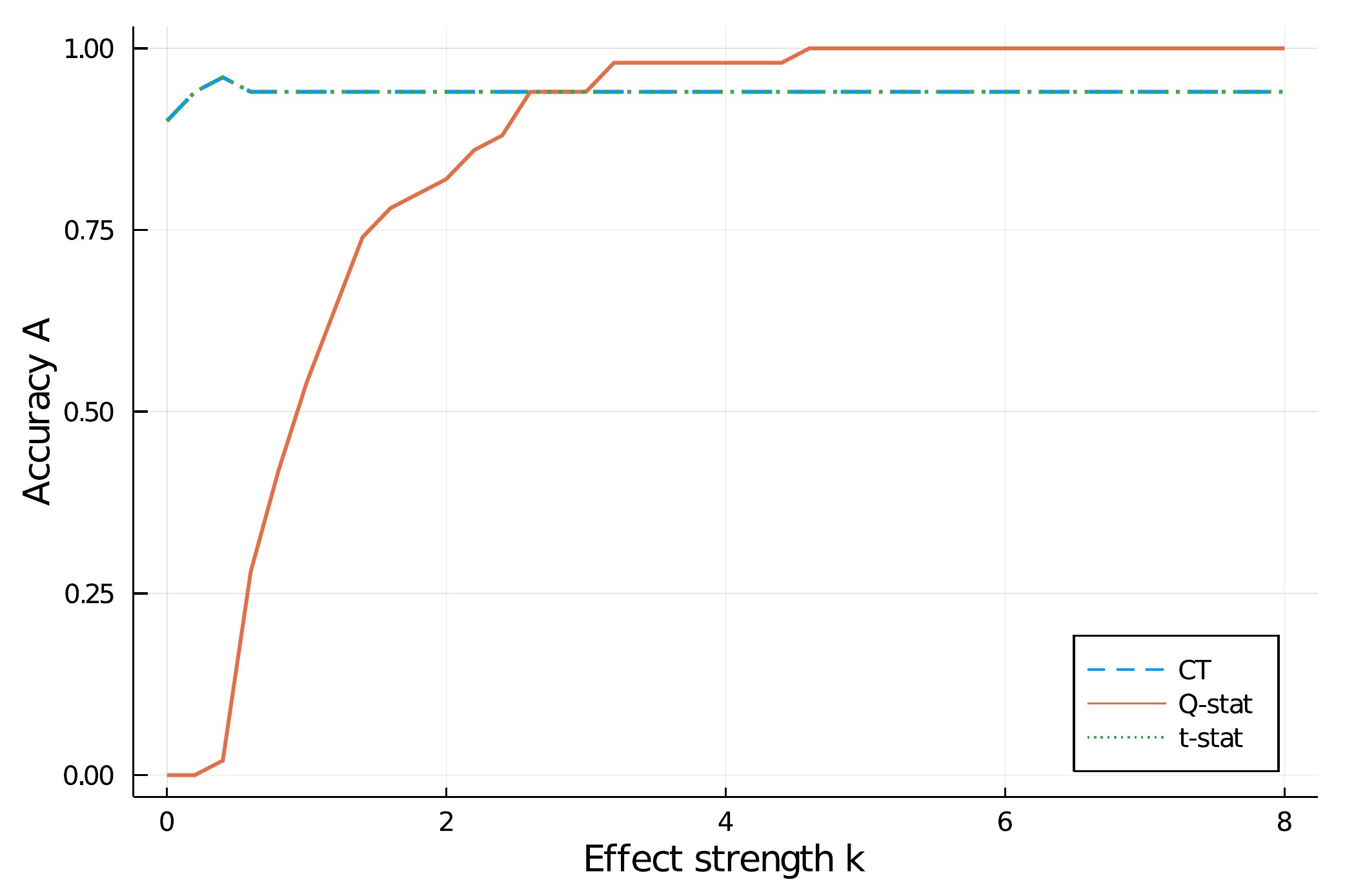}
    \end{subfigure} %
    \begin{subfigure}{.45\linewidth}
    \includegraphics[width=\linewidth]{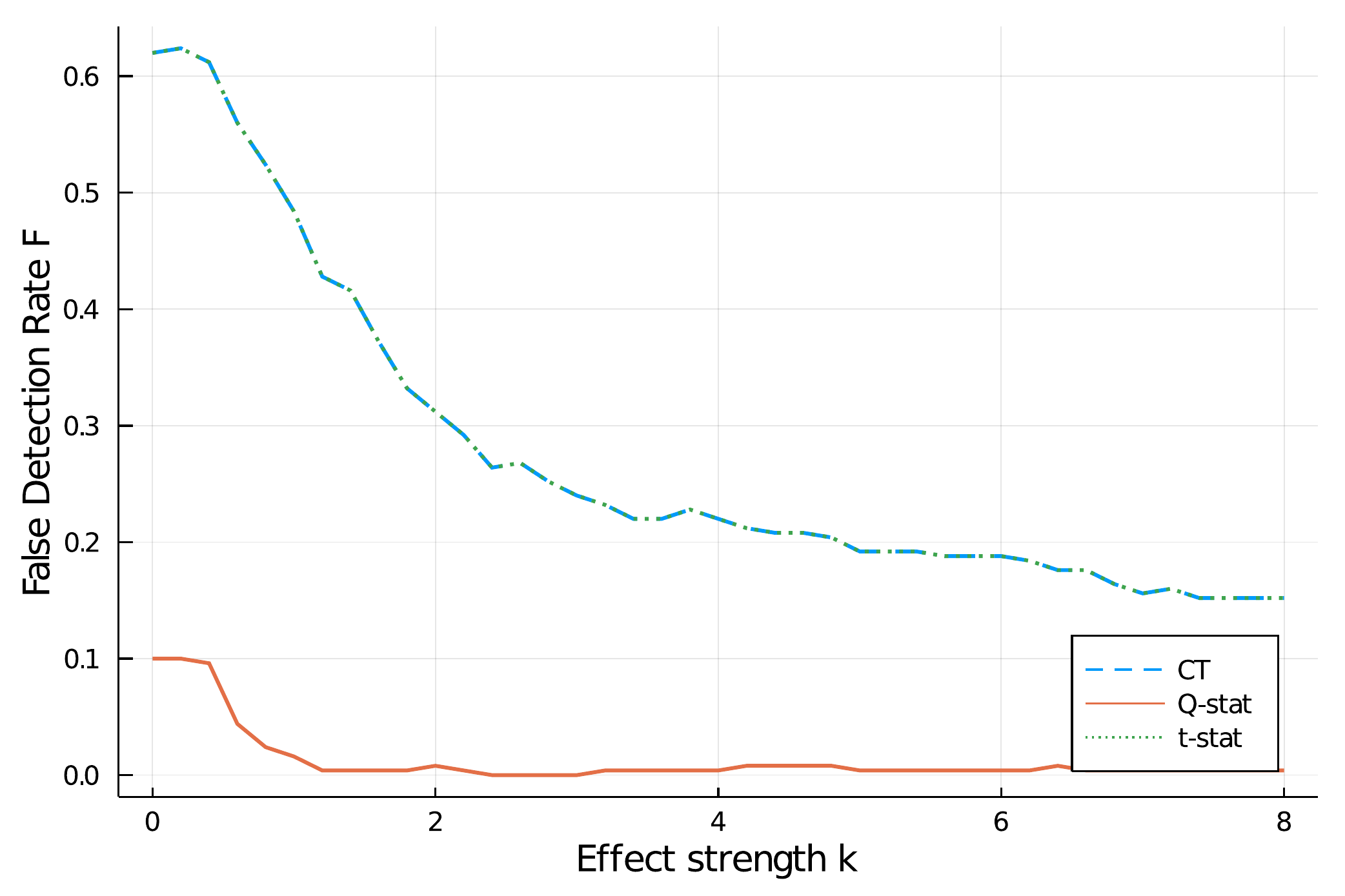}
    \end{subfigure}
    \caption{Accuracy (left panel) and False detection rate (right panel) as the effect strength $k$ increases, to measure the ability of each HTE method to isolate features affecting treatment heterogeneity. Trees obtained by ``$t$-test'' and ``CT'' are pruned.  Metrics are averaged over $20$ random synthetic datasets with 2-rule, containing $N=1,000$ observations each.}
    \label{fig:syn.ortho}
\end{figure}

To evaluate the robustness of our recursive partitioning procedure to the confounding factors, we conduct a similar set of experiments except that the treatment effect is now defined as $\tau_k(X_4,X_5)$. In other words, $X_1, X_2, X_3$ impact the baseline effect $Y(0)$ and the propensity score $\mathbb{P}(T=1|\bm{X}=\bm{x})$, while $X_4,X_5$ drive the treatment effect heterogeneity. If $\mathcal{S}$ denotes the set of features the tree splits on, we evaluate the accuracy $A = |\mathcal{S}\cap \{4,5\}|/2$ and false detection rate $F = |\mathcal{S} \setminus \{4,5\}|/|\mathcal{S}|$. With these notations, a good HTE algorithm should split on $X_4$ and $X_5$, i.e., $A=1$, and $X_4,X_5$ only, i.e., $F=0$. Figure \ref{fig:syn.ortho} compares the accuracy (left panel) and false detection rate (right panel) of all three methods as the effect strength $k$ increases, for the 2-rule treatment. Causal trees and $t$-test based trees are pruned. For each metric, the cross-validation criterion for these two methods is chosen so as to optimize this particular metric. Figure \ref{fig:syn.ortho} suggests that our recursive partitioning procedure is more robust to confounding than either causal trees or $t$-test based tests.

\FloatBarrier
\section{Supporting material for the revenue management example} \label{sec:A.jdcom}
This section provides details on the analysis of the JD.com data \citep{shen2019jd}.

\subsection{Data processing} \label{ssec:A.jdcom.process}
We aggregated data from different sources, which we briefly present in this section. We emphasize the data characteristics that are relevant to our analysis but refer to \citet{shen2019jd} for a more detailed presentation of the data. 

{\it Clicks data: } This data contains the list of pages/products visited by each user, the channel they used (e.g., desktop, mobile, app), and the corresponding time-stamp. We say that two pages are part of the same session if they are visited by the same user, on the same platform, within $10$ minutes. If a user visited the same page multiple times within the same session, we only retain her last visit.

{\it Users data: } Each user is described by a unique identifier, some statistics on her purchase history (number of past purchases, time of first purchase, loyalty program), and demographics (e.g., gender, education). 

{\it Products data: } We characterize each product by its type and two attributes. Given the large proportion of products with missing attributes, we only use the presence of each attribute as a covariate, not its value (when available). With deanonymized data, product category-specific imputations could be considered as well.

{\it Orders data: } For each order, we have access to the initial and final price, the difference indicating whether a discount was applied. In our analysis, we do not discriminate between the different types of discounts (direct, quantity, bundle,...). For unbought items, we do know neither the original price nor whether the price was discounted. For the original price, we impute it using the closest (in time) price for the same product in the training data - the ``original price'' listed in the data is product-specific and the same for all users. For the discount indicator, we apply our robust estimation strategy from Section \ref{sec:robust}. We exclude from our analysis products that have never been purchased, products whose price changed more than $62$ times (twice a day) over the month of March 2018, and transactions consisting of free products. 

All together, the final dataset consists of $566,004$ customer-product interactions, described with 9 variables concerning the user (number of past purchase, time since first purchase, ``Plus'' membership indicator, age, gender, education, marital status, purchase power category, city level), 4 variables concerning the product (product type, original price, attribute 1 and 2 missingness indicator), and 2 variables regarding the interaction (day of the week, hour of the day).

\subsection{Predicting discount assignment} \label{ssec:A.jdcom.ml}
A predictive model to predict discount assignment can be used to understand the drivers of discounts (or back-engineer the promotion targeting strategy). Table \ref{tab:drivers} reports the most important features for the decision tree and random forest models. In both cases, original price is the most important feature, with a clear edge over the second best one. 
\begin{table}[h]
\footnotesize
\caption{Key parameters driving the application of a promotion on bought items. For each variable, we also report the importance value, according to each model.}
\label{tab:drivers}
\centering
\begin{tabular}{cll}
Variable rank & Decision tree & Random forest \\ 
\toprule
1 & Original price (0.59) & Original price (0.25)  \\
2 & Product type (0.08) & Time since 1st purchase (0.15)  \\
3 & Time since 1st purchase (0.06) & Hour of the day (0.14) \\
4 & Attribute 2 missing (0.03) & City type (0.06)  \\
5 & Hour of the day (0.03) & Number of past purchases (0.05) \\
\bottomrule
\end{tabular}
\end{table}

\subsection{Heterogeneous odds ratio estimation} \label{ssec:A.jdcom.hor}
We first consider the vector of treatment assignments $\bm{t}$ leading to the worst odds ratio, i.e., the closest to $1$. By doing so, we obtain the following the $2\times 2$ contingency table:
\begin{table}[h]
\footnotesize
\caption{$2\times 2$ contingency table on JD.com's data with adversarial treatment assignments.}
\label{tab:cont.jd}
\centering
\begin{tabular}{c|cc}
    & $Y=0$ & $Y=1$  \\
    \midrule
    $T=0$ &  $218,069$ & $16,399$\\
    $T=1$ & $305,751$ & $25,785$
\end{tabular}
\end{table}


\end{document}